\documentclass[12pt]{article}
\pdfoutput=1

\usepackage{amsmath,amssymb,amsbsy,amsfonts,latexsym,graphicx}
\usepackage{color,array,subfigure}
\usepackage{cite}

\newcommand{\cals}{\mbox{${\cal S}$}}

\newcommand{\nn}{\nonumber}

\newcommand{\tr}{{\rm tr}\,}

\allowdisplaybreaks

\evensidemargin \oddsidemargin

\def\nn{\nonumber}

\begin{document}


\begin{titlepage}

\renewcommand{\thefootnote}{\fnsymbol{footnote}}


\begin{flushright}
\end{flushright}

\vspace{7mm}
\baselineskip 7mm
\begin{center}
  {\Large \bf AdS Q-Soliton and Inhomogeneously mass-deformed ABJM Model}
\end{center}

\begin{center}
\vskip 0.5cm
  {Byoungjoon Ahn$^{a}$\footnote{e-mail : bjahn@yonsei.ac.kr}, Seungjoon Hyun$^{a}$\footnote{e-mail : sjhyun@yonsei.ac.kr}, Kyung Kiu Kim$^{b}$\footnote{e-mail : kimkyungkiu@sejong.ac.kr}, O-Kab Kwon$^{c}$\footnote{e-mail : okab@skku.edu} \\and Sang-A Park$^{a,d}$\footnote{e-mail : sapark@kias.re.kr}}

\vskip 0.3cm

{\it $^a\,$ Department of Physics, Yonsei University, Seoul 03722, Korea }\\
{\it $^b\,$ Department of Physics and Astronomy, Sejong University, Seoul 05006, Korea }\\
{\it $^c\,$ Department of Physics, Autonomous Institute of Natural Science, Institute of Basic Science,
Sungkyunkwan University, Suwon 16419, Korea } \\
{\it $^d\,$ School of Physics, Korea Institute for Advanced Study, Seoul 02455, Korea } \\

\end{center}

\thispagestyle{empty}

\vfill
\begin{center}
{\bf Abstract}
\end{center}
\noindent

We study dual geometries to a deformed ABJM model with spatially dependent source functions at finite temperature. These source functions are proportional to the mass function $m(x)= m_0 \sin k x$ and its derivative $m'(x)$. As dual geometries, we find hairy black branes and AdS solitons corresponding to deconfinement phase and confining phase of the dual field theory, respectively. It turns out that the hairy AdS solitons have lower free energy than the black branes when the Hawking temperature is smaller than the confining scale. Therefore the dual system undergoes the first order phase transition. Even though our study is limited to the so-called Q-lattice ansatz, the solution space contains a set of solutions dual to a supersymmetric mass deformation. As a physical quantity to probe the confining phase, we investigate  the holographic entanglement entropy and discuss its behavior in terms of modulation effect.
\\ [13mm]

\vspace{5mm}
\end{titlepage}

\baselineskip 6.6mm
\renewcommand{\thefootnote}{\arabic{footnote}}
\setcounter{footnote}{0}

\section{Introduction}

The ${\cal N} = 8$ SO(8) gauged supergravity in 4-dimensions~\cite{deWit:1981sst} can be obtained by the consistent KK truncation of 11-dimensional supergravity on $S^7$. The resulting theory can be truncated further to ${\cal N} = 4$ SO(4) gauged supergravity~\cite{Das:1977pu}. Our starting action is the bosonic sector of this ${\cal N} =4$ theory coupled to a single chiral multiplet with vanishing gauge fields\footnote{This action can also be obtained as a truncation of the ${\cal N} = 2$ STU gauged supergravity theory~\cite{Cvetic:1999xp,Cvetic:2000tb}. }, which can also be the bosonic sector of the ${\cal N} = 1$ supergravity theory in 4-dimensions.  
For this reason, any solution of the action in this paper belongs to SO(4)$\times$SO(4) invariant sector of the ${\cal N} = 8$ gauged supergravity theory and can be a uplifted solution with SO(4)$\times$SO(4) isometry in the 11-dimensional supergravity~\cite{Cvetic:1999au}.

The isometry SO(4)$\times$SO(4) in the 11-dimensional supergravity can be extended to the case of SO(4)/$\mathbb{Z}_q\times {\rm SO(4)}/\mathbb{Z}_q$, which corresponds to  the ${\cal N} = 6$ U$_q(N)\times {\rm U}_{-q}(N)$ ABJM theory~\cite{Aharony:2008ug} at the Chern-Simons level $q$. This ABJM theory also admits a supersymmetry preserving  deformation with {\it constant} mass (mABJM)~\cite{Hosomichi:2008jb,Gomis:2008vc}. The gravity dual of the mABJM theory is known as the LLM geometries~\cite{Lin:2004nb} with $\mathbb{Z}_q$ orbifold. The duality of LLM/mABJM  using the holographic renormalization method was investigated in \cite{Jang:2016aug,Jang:2019pve}.

The supersymmetric Q-lattice (Susy Q) solution preserving  1/4 supersymmetry was studied in \cite{Gauntlett:2018vhk}. This geometry is regarded as the gravity dual of an ${\cal N} = 3$ inhomogeneously mass-formed ABJM (ImABJM) 
model~\cite{Kim:2018qle,Kim:2019kns} with a certain periodic mass function at the zero temperature ($T=0$).\footnote{ The dual geometries of the ImABJM models also have the same isometry with those of the mABJM models, except for the worldvolume direction of M2-branes.} 
Various aspects of the ImABJM models including lower supersymmetric models and vacuum solutions for periodic mass function were investigated in \cite{Kim:2019kns}. 
One of interesting properties of the Susy Q solution is the fact that the solution describes the boomerang RG flow from the AdS geometry in the UV to the same AdS geometry in the 
IR~\cite{Donos:2017ljs,Donos:2017sba,Chesler:2013qla}. On the other hand, since the mass function is originated from the 4-form field strength with the dependence of a worldvolum coordinate of M2-branes~\cite{Kim:2018qle} and the Susy Q solution has a special form of the mass function, one may take an arbitrary function as the mass function extending the Susy Q solution. There are other dual solutions~\cite{DHoker:2009lky,Bobev:2009ms,Bobev:2013yra,Arav:2018njv} to the ImABJM models, which are not included in the Susy Q solution.

However, no work has been done about the ImABJM models in the finite temperature ($T>0$) so far. Therefore we focus on the gravity dual to a finite temperature ImABJM model with a special choice of mass function in this work. A gravity dual to a finite temperature system may be given by a black brane. Thus it is very natural to find a black brane solution admitted by the action (\ref{bulk_action}) studied in  \cite{Gauntlett:2018vhk,Arav:2018njv}. The action in question contains a complex scalar field dual to the conformal dimension 1 and 2 operators in the ABJM theory. The sources corresponding to these operators are proportional to the mass function $m(x)$ and its derivative $m'(x)$, which appear in the ImABJM models~\cite{Kim:2018qle}. In this paper, we consider a sinusoidal form of the mass function. So the scalar field and the metric can be taken as the Q-lattice type ansatz (\ref{BBrane00}), which has been used to study  momentum relaxation in the context of the AdS/CMT, {\it e.g.} \cite{Donos:2013eha,Ling:2014laa,Kim:2016jjk,Ahn:2017kvc}. We provide the numerical solutions of the black branes and the boundary quantities based on the numerical solutions using holographic renormalization. The detailed study on the thermodynamics of these Q-lattice black branes will be reported soon \cite{Hyun:2019juj}.

On the other hand, there is another kind of solutions competing with these black branes. The action (\ref{bulk_action}) also admits a horizon-less configuration (\ref{solition01}) that is obtained by a double Wick rotation from the black brane geometry. When the scalar field vanishes, the background geometry is known as the AdS soliton \cite{Horowitz:1998ha,Nishioka:2006gr,Nishioka:2009zj}. This background has a tip accompanied by a circle as a regular geometry. The period of the circle describes the confining scale since the AdS soliton is regarded as a dual geometry of the confining phase. Then, turning on the spatially dependent mass function, the dual geometry becomes a solitonic geometry with a scalar hair. Like the hairless case, this geometry is also obtained by the double Wick rotation from the Q-lattice black brane solution. So we call this solitonic geometry ``AdS Q-soliton'' from now on. It turns out that this AdS Q-soliton dominates the system in the low temperature. We confirm this fact by comparing the free energy which is equivalent to the gravity on-shell action. Therefore, the gravity system (\ref{bulk_action}) admits the hairy black brane and the AdS Q-soliton, which are preferable in high temperature and low temperature, respectively. The phase boundary only depends on temperature.

Correspondingly, the AdS Q-soliton is more important in low temperature and it implies that there is a confining phase governing the low temperature physics. One of physical quantities which can probe confinement is known as the holographic entanglement entropy (HEE) \cite{Klebanov:2007ws,Nishioka:2006gr}. This quantity distinguishes the phases by the entanglement entropy curve under varying the size of entangling region. The curve of the confining phase has a characteristic length $l_{\text{crit}}$ which could be of importance to understand the confinement. 
We pick  a solution set dual to the ImABJM model in the entire solution space and study the modulation effect for $l_{\text crit}$ in the selected AdS Q-soliton backgrounds. This analysis shows that the modulation reduces the characteristic length and enhances the confinement.

This paper is organized as follows. In section 2, we introduce the gravity model and provide the numerical solutions with evaluating various physical quantities. In section 3, we construct the phase diagram of the gravity model and find the set of solutions corresponding to the ImABJM model with $m(x)= m_0 \sin k x$. In section 4, we study the HEE in AdS Q-solitons using a numerical method and we find modulation effect  of the characteristic length $l_{\text{crit}}$. In section 5 we conclude by discussion and future directions.

\section{Q-lattice Black Brane and AdS Q-Soliton}\label{BBG}

In this section, we find  gravity solutions which are dual to a deformed ABJM theory with spatially dependent source functions $\mathcal{J}_{\mathcal X}(x)$ and $\mathcal{J}_{\mathcal Y}(x)$, where $x$ is one of  spatial coordinates. Corresponding operators $\mathcal{O}_{\mathcal X}(x)$ and $\mathcal{O}_{\mathcal Y}(x)$ also depend on the coordinate $x$. When we consider a particular case with 
\begin{align}
\left\{\mathcal{J}_{\mathcal X},\mathcal{J}_{\mathcal Y}\right\}=\left\{ m'(x), m(x) \right\}\nonumber
\end{align} 
and 
\begin{align}\label{SusyDeformation}
\left\{\mathcal{O}_{\mathcal X},\mathcal{O}_{\mathcal Y} \right\} = \left\{ M_A^B  \tr \left(Y^A Y_B^\dagger\right) ,\, M_A^B \tr\left( \psi^{\dagger A}\psi_B + \frac{8\pi}{q} Y^C Y^\dagger_{[C} Y^A Y^\dagger_{B]} \right) \right\},
\end{align}
the deformations preserve $\mathcal{N}=3$ supersymmetry\footnote{There is one more operator to complete supersymmetry transformation but the operator doesn't affect the dual geometry. See \cite{Gauntlett:2018vhk}.} \cite{Kim:2018qle,Kim:2019kns}. Here $q$ denotes the Chern-Simons level and $M_A^B$ is the component of a diagonal matrix given by $diag(1,1,-1,-1)$. So the indices are running from 1 to 4. $Y^A$ and $\psi^A$ are the scalar field and the fermion field in the ABJM theory. See \cite{Kim:2019kns} for details.  In this work we study gravity dual to the deformation in a finite temperature. The corresponding gravity model at zero temperature is studied in \cite{Gauntlett:2018vhk,Arav:2018njv} via gauge/gravity correspondence. Therefore we start with the bosonic part of the supergravity action introduced in \cite{Gauntlett:2018vhk,Arav:2018njv} as follows: 
\begin{equation}\label{bulk_action}
\cals_B=\frac{1}{16\pi G}\int d^4x \sqrt{-g} \big( R - \frac{2}{(1-|z|^2)^2}\partial_M z \, \partial^M \bar{z} +\frac{1}{L^2}\frac{2(3-|z|^2)}{1-|z|^2}\big)~,
\end{equation}
where the complex scalar field $z=\mathcal{X}+ i\mathcal{Y}$ is dual to $\mathcal{O}_\mathcal{X}$ and $\mathcal{O}_\mathcal{Y}$. The sources and the operators in (\ref{SusyDeformation}) are related to the asymptotic behavior of the bulk field $z$ up to normalization as follows:
\begin{align}
z(r,x) \sim \frac{\mathcal{O}_{\mathcal{X}}(x) + i \mathcal{J}_{\mathcal{Y}}(x)}{r} + \frac{\mathcal{J}_{\mathcal{X}}(x)) + i \mathcal{O}_{\mathcal{Y}}(x)}{r^2} + \ldots ,
\end{align}
where $r$ is the radial coordinate and the boundary of the AdS space is located at $r=\infty$.

For black branes, we impose the following ansatz:
\begin{align}\label{BBrane00}
&ds^2 = -\frac{U(r)}{L^2} e^{2W_0(r) } dt^2 +\frac{r^2}{L^2} e^{2W_1(r) } dx^2 +\frac{r^2}{L^2}   dy^2 + \frac{ L^2 dr^2}{U(r)}~,
\nonumber\\
&z(r,x) = \mathcal{R}(r) e^{i k x}~.
\end{align}
This configuration is called Q-lattice which is useful to see the modulation effect in holographic study \cite{Donos:2013eha}. Taking this ansatz, the sources and the vacuum expectation values(VEVs) of the oprators in (\ref{SusyDeformation}) become trigonometric functions of $x$.

\begin{figure}[ht!]
\centering
    \subfigure[ ]
    {\includegraphics[width=7.5cm]{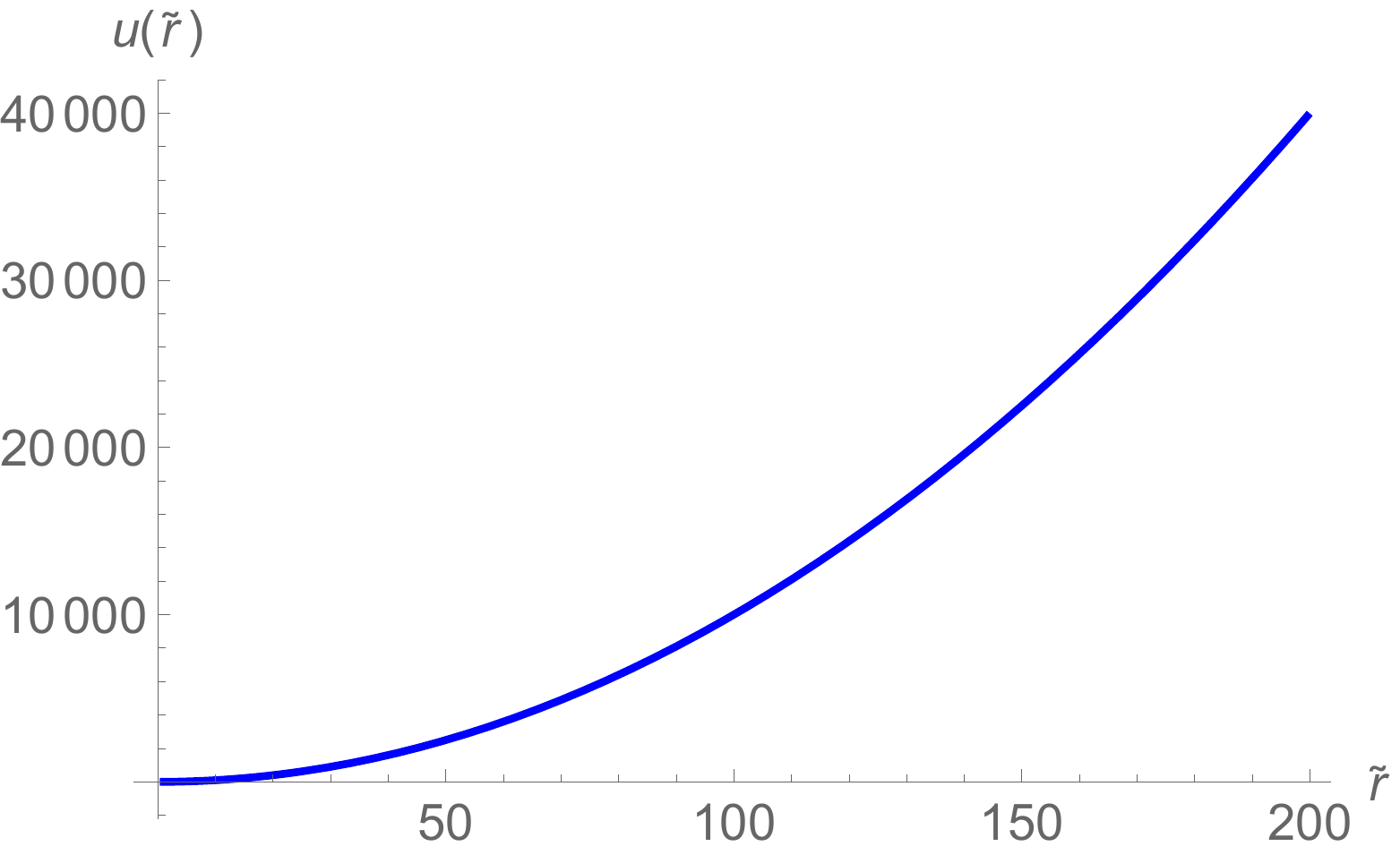}  }
\subfigure[ ]
{\includegraphics[width=7.5cm]{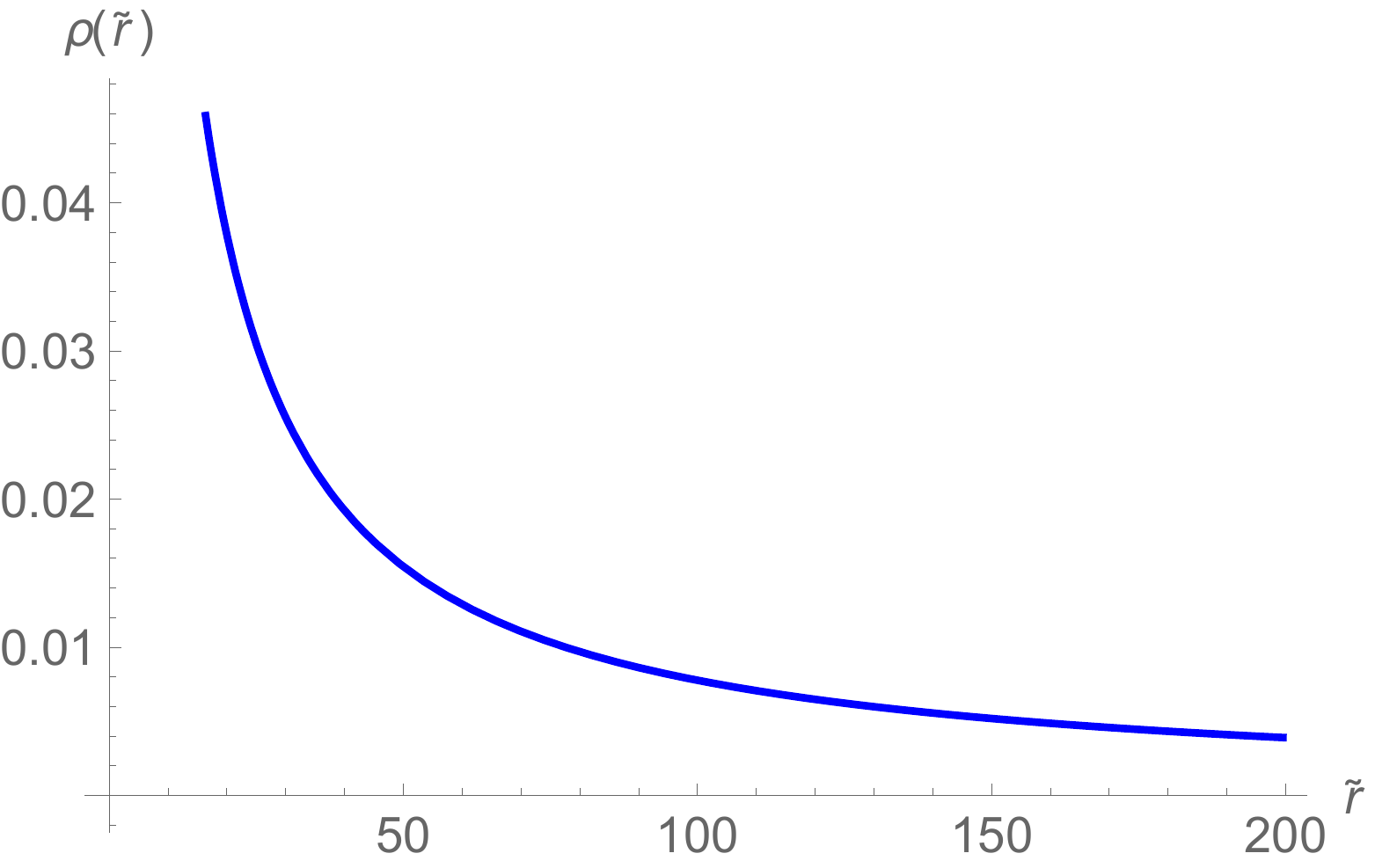}  }

    \caption{{\bf Numetrical solutions for a Q-lattice Black brane and a AdS Q-soliton with $\tilde{k}=0.6336$, $\rho(1)=0.4000$ and $w_0(1)=w_1(1)=10^{-4}$}
} \label{fig:HBH01}
\end{figure}

The equations of motion are given by second order differential equations for $W_1(r)$ and $\mathcal{R}(r)$ and first order differential equations for $U(r)$ and $W_0(r)$. Furthermore we take the following scaling:
\begin{align}\label{scaling00}
&r = r_h \tilde{r}~,~U(r) = r_h^2 u(\tilde r)~,~\mathcal{R}(r) = \rho(\tilde r)~,\nonumber\\
&W_0(r)= w_0(\tilde r)~,~W_1(r)= w_1(\tilde r)~,~k = \frac{r_h}{L^2} \tilde{k}~,
\end{align}
where $r_h$ is the location of the horizon. Then the regularity conditions at the horizon is given as follows:
\begin{align}\label{regul_con_BH}
&u(1)=0~,~\nonumber\\
&\rho '(1)=-\frac{\rho(1) e^{-2 w_1(1)} \left(\tilde{k}^2 \left(\rho(1)^2+1\right)+2 \left(\rho(1)^2-1\right) e^{2 w_1(1)}\right)}{\rho(1)^2-3}~,\nonumber\\
&w_1'(1) = -\frac{2 \tilde{k}^2 \rho(1)^2 e^{-2 w_1(1)}}{\rho(1)^4-4 \rho(1)^2+3}~.
\end{align}
The detailed explanation to obtain these conditions is provided in Appendix B. Therefore, a hairy black brane solution is parametrized by $r_h$, $\rho(1)$, $w_0(1)$, $w_1(1)$ and $\tilde{k}$. However $w_0(1)$ and $w_1(1)$ do not play any crucial role in this system. A choice of these values changes $w_0(\infty)$ and $w_1(\infty)$ which are compensated by a trivial coordinate transformation:
\begin{align}
t \to e^{W_0(\infty)}t~,~x \to e^{W_1(\infty)}x ~,~ k\to e^{-W_1(\infty)} k.
\end{align}
Thus $w_0(1)$ and $w_1(1)$ can be fixed by suitable values and the relevant parameters of the black brane are $r_h$, $\rho(1)$ and $\tilde{k}$. We plot a numerical solution in Figure \ref{fig:HBH01}. For $k= 0$ case, the numerical solution is reduced to that in \cite{Hertog:2004ns}.

The temperature and entropy density of the black branes are given by
\begin{align}\label{Temp}
T =& \frac{U'(r_h)}{4\pi L^2}e^{W_0\left(r_h\right)-W_0(\infty )}
 =  \frac{r_h}{4\pi L^2} u'(1) e^{w_0\left(1\right)-w_0(\infty )}= \frac{r_h}{4\pi L^2}e^{w_0\left(1\right)-w_0(\infty )}\frac{3-\rho(1)^2}{1-\rho(1)^2}~,~ \\
s =& \frac{ r_h^2}{4 G L^2}  e^{W_1(r_h)-W_1(\infty)}= \frac{ r_h^2}{4 GL^2}  e^{w_1(1)-w_1(\infty)}~.
\end{align}
On the other hand, we need holographic renormalization \cite{Balasubramanian:1999re,deHaro:2000vlm,Skenderis:2002wp} to evaluate various physical quantities in the dual field theory. See \cite{Cabo-Bizet:2017xdr,Gauntlett:2018vhk} for closer applications to this work. Such quantities are determined by the asymptotic behavior of the fields. One can find the asymptotic behavior by solving the equations of motion in the large $r$ expansion. The asymptotic expansion of the fields is given by 
\begin{align}\label{AsympExp}
&u(\tilde r) \sim~  \tilde{r}^2+\tilde{\rho }_1^2-\frac{\tilde{\mathit{m}}}{\tilde{r}}+\frac{2 \tilde{\rho }_2^2}{\tilde{r}^2}+ \frac{ 8 \tilde{k}^2\tilde{\rho }_1 \tilde{\rho }_2 e^{-2 \tilde{w}_{1,0}}-\tilde{\mathit{m}} \tilde{\rho }_1^2 }{6 \tilde{r}^3}~, 
\nonumber\\
&w_0(\tilde r) \sim ~ \tilde{w}_{0,0} -\frac{\tilde{\rho }_1^2}{2 \tilde{r}^2}+\frac{-\frac{4}{3} \tilde{\rho }_1 \tilde{\rho }_2-\tilde{w}_{1,3}}{\tilde{r}^3}+\frac{\frac{\tilde{\rho }_1^4}{4}-\tilde{\rho }_2^2}{\tilde{r}^4}+ \cdots~, 
\nonumber\\
&w_1(\tilde r)\sim ~ \tilde{w}_{1,0}+\frac{\tilde{w}_{1,3}}{\tilde{r}^3}-\frac{\tilde{k}^2 \tilde{\rho }_1^2 e^{-2 \tilde{w}_{1,0}}}{2 \tilde{r}^4}+\cdots~, 
\nonumber\\
&\rho(\tilde r)\sim ~  \frac{\tilde{\rho }_1}{\tilde{r}}+\frac{\tilde{\rho }_2}{\tilde{r}^2}+\frac{\tilde{\rho }_1 \left(\tilde{k}^2 e^{-2 \tilde{w}_{1,0}}-\tilde{\rho }_1^2\right)}{2 \tilde{r}^3}
 +\cdots~.
\end{align}
Also, one can easily see that the ansatz and this asymptotic behavior admit the usual 
AdS-Schwarzschild black brane solution when the scalar field vanishes:
\begin{align}
U(r) = r^2 - \frac{m}{r} ~,~~W_1(r)=W_2(r)=0~,~~\mathcal{R}(r)=0~.
\end{align}

Now, one may consider a horizon-less solution of (\ref{bulk_action}). Such a solution is nothing but a hairy AdS Soliton solution whose metric is obtained by a double Wick rotation from the black brane metric. The form of metric and scalar field is given by
\begin{align}\label{solition01}
&ds^2 = \frac{U(r)}{L^2} e^{2W_0(r)} d\chi^2 +\frac{r^2}{L^2} e^{2W_1(r)} dx^2 -\frac{r^2}{L^2}  dt^2 + \frac{ L^2 dr^2}{U(r)}~, 
\nonumber\\
&z(r) = \mathcal{R}(r) e^{i k x}~,
\end{align} 
where $\chi$ is a periodic coordinate. The period of the circle is taken as
\begin{align}
\chi_p = \frac{4\pi L^2}{U'(r_0)} e^{-(W_0(r_0)-W_0(\infty))}~
\end{align}
to make the geometry regular at the tip location, $r=r_{0}$. Here we take the boundary metric as $\eta_{\mu\nu}=diag(-1,1,1)$. The equations of motion for this hairy AdS Soliton also admit the same scaling of (\ref{scaling00}) with $r_0$ replacing $r_h$.  Taking into account this scaling, the equations of motion for $u$, $w_0$, $w_1$ and $\rho$ become the same expressions as those of the hairy black brane. Thus our numerical solution can be identified with the black brane as well as the AdS Soliton with the complex scalar hair. Of course, the asymptotic expansion (\ref{AsympExp}) is also valid for the hairy AdS Soliton by changing $\tilde r = r/r_0$. Since the AdS Soliton geometry usually describes a confining phase, one important quantity is the confining scale given by the inverse period of the $\chi$-circle:
\begin{align}\label{ConScale}
\Lambda_{0} = \frac{r_0}{4\pi L^2}e^{w_0\left(1\right)-w_0(\infty )}\frac{3-\rho(1)^2}{1-\rho(1)^2}~.
\end{align}
This is the same expression as the temperature (\ref{Temp}) after replacing  $r_h$ by $r_0$. As we mentioned in the introduction, we call this solution AdS Q-soliton.

\section{Phase Diagram}\label{HRN}

In this section we study the phase diagram which consists of the Q-lattice black brane phase and the AdS Q-soliton phase by comparing the free energy. They correspond to the deconfinement phase and the confining phase, respectively. As a physical quantity distinguishing two phases, we will investigate the entanglement entropy for the confining phase in the next section.

In order to compare two phases, firstly we need to identify physical parameters of the phase space in terms of numerical quantities. We hope that the gravity solutions in the previous section can be dual to the ImABJM theory with a mass function $m(x) = m_0 \sin k x$. Therefore desirable physical parameters\footnote{Here all the physical quantities are normalized with the boundary metric $\eta_{\mu\nu}$.} of the system are $m_0$, $k$ and the confining scale $\Lambda_0$ or the temperature $T$. To identify $m_0$ using quantities from the geometry, we may consider Ward identities. In this system we have two Ward identity as follows:
\begin{align}
\partial_\mu \left< \mathcal{T}^{\mu\nu} \right>  =&  \left<\mathcal{O}_{\mathcal{X}}\right> \partial^\nu \mathcal{J}_{\mathcal{X}}+  \left<\mathcal{O}_{\mathcal{Y}}\right> \partial^\nu \mathcal{J}_{\mathcal{Y}} \\
\left< {\mathcal{T}^\mu}_\mu \right> =& \left(3-\Delta_{\mathcal{X}}\right)\left<\mathcal{O}_{\mathcal{X}}\right>   \mathcal{J}_{\mathcal{X}}+\left(3-\Delta_{\mathcal{Y}}\right)  \left<\mathcal{O}_{\mathcal{Y}}\right>   \mathcal{J}_{\mathcal{Y}}~.
\end{align}
From the VEVs given by the holographic renormalization in Appendix A, one can easily check that the Q-lattice black brane and the AdS Q-soliton satisfy the above Ward identities by choosing sources as follows:
\begin{align}
\mathcal{J}_{\mathcal{X}} = -\frac{\tilde{\rho}_2 r_{h,0}^2}{L^4} \cos {k}{x}=j_{\mathcal X}\cos {k}{x}~,~~\mathcal{J}_{\mathcal{Y}} = \frac{\tilde{\rho}_1 r_{h,0}}{L^2} \sin {k}{x}=j_{\mathcal Y} \sin {k}{x}~,
\end{align}
where the subscripts $h$ and $0$ stand for the location of the horizon in the black brane geometry and the location of tip in the AdS soliton geometry, respectively. These sources can be related to the mass function of the dual field theory.

Now let us fix $m_0$ in terms of geometric parameters as follows\footnote{There is a freedom to identify a source in terms of the asymptotic value of the scalar field. This freedom can not be determined by the Ward identities or other physical relations. Thus one can fix one source in terms the geometric data using this freedom.}:
\begin{align}\label{m0}
m_0 =j_{\mathcal Y}= \frac{\tilde{\rho}_1 r_{0}}{L^2} = \frac{\tilde{\rho}_1 r_{h}}{L^2}~.
\end{align}
Then the other source $j_{\mathcal X}$ can be written as:
\begin{align}\label{eta_k}
m_0 k  \,\eta  = j_{\mathcal X} = - \frac{\tilde{\rho}_2 r_{0}^2}{L^4} = -\frac{\tilde{\rho}_2 r_{h}^2}{L^4}~,
\end{align}
where $\eta$ can be read off from numerical solutions. In general $\eta$ is not equal to $1$. 
In this work, we limit our study on geometries with the Q-lattice ansatz and the sources proportional to $m(x)$ and $m'(x)$. So we also take into account solutions with $\eta\neq 1$.\footnote{We  speculate that $\eta$ could be absorbed to normalization of the operators or could describe more general deformation. Since we need more careful study including general ansatz beyond the Q-lattice, we just keep $\eta$ as a parameter in the following formulations. We leave the general case as a future work.}   
Another physical parameter specifying the dual system is the modulation number $k$ that is related to the dimensionless modulation $\tilde{k}$ as follows:
\begin{align}\label{kphy}
k = e^{-w_1(\infty)} \frac{r_h}{L^2}\tilde{k} = e^{-w_1(\infty)} \frac{r_0}{L^2}\tilde{k}~.
\end{align}

In summary the physical parameters specifying the system are $k$, $m_0$ and $T$ or $\Lambda_0$. These parameters are given by $\tilde k$, $\rho(1)$, $w_0(1)$, $w_1(1)$ and $r_h$ or $r_0$. Since $w_0(1)$ and $w_1(1)$ do not change the geometry meaningfully as we discussed earlier, one can always take certain values for $w_0(1)$ and $w_1(1)$. Therefore the relevant physical parameters for this system are $\tilde k$ and $\rho(1)$ or $\tilde{\rho}_1$ including  $r_h$ or $r_0$. Furthermore, $r_h$ and $r_0$ are not physical parameters in the dual field theory. Rather, the temperature (\ref{Temp}) and the confining scale (\ref{ConScale}) of the system are physically relevant parameters. So the deconfinement phase (Q-lattice Black brane) and the confining phase (AdS Q-Soliton) are parametrized by ($T$, $m_0$, $k$) and ($\Lambda_0$, $m_0$, $k$), respectively. 

\begin{figure}[ht!]
\centering
    \subfigure[ ]
    {\includegraphics[width=6.7cm]{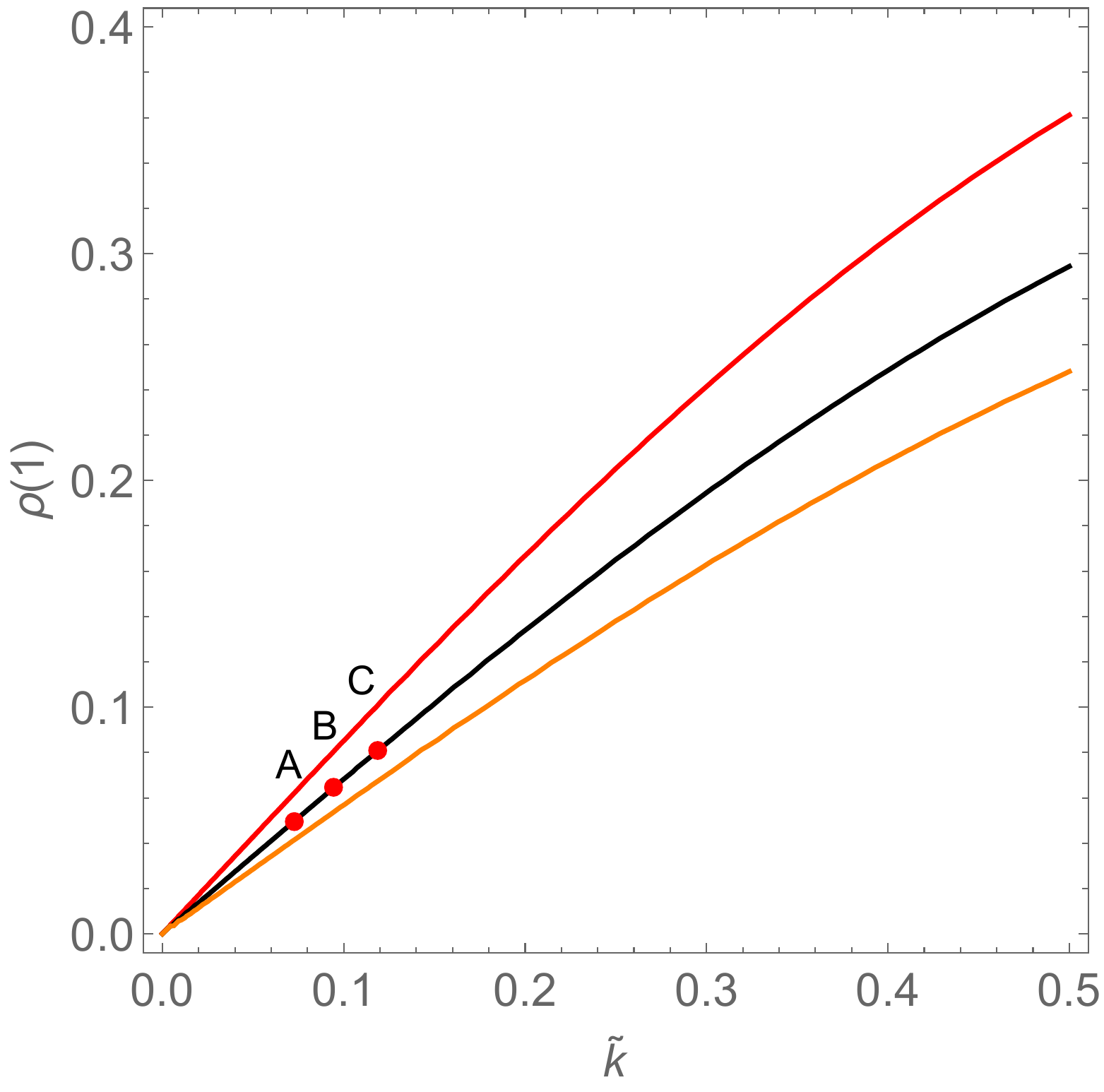}  }
\subfigure[ ]
{\includegraphics[width=8.3cm]{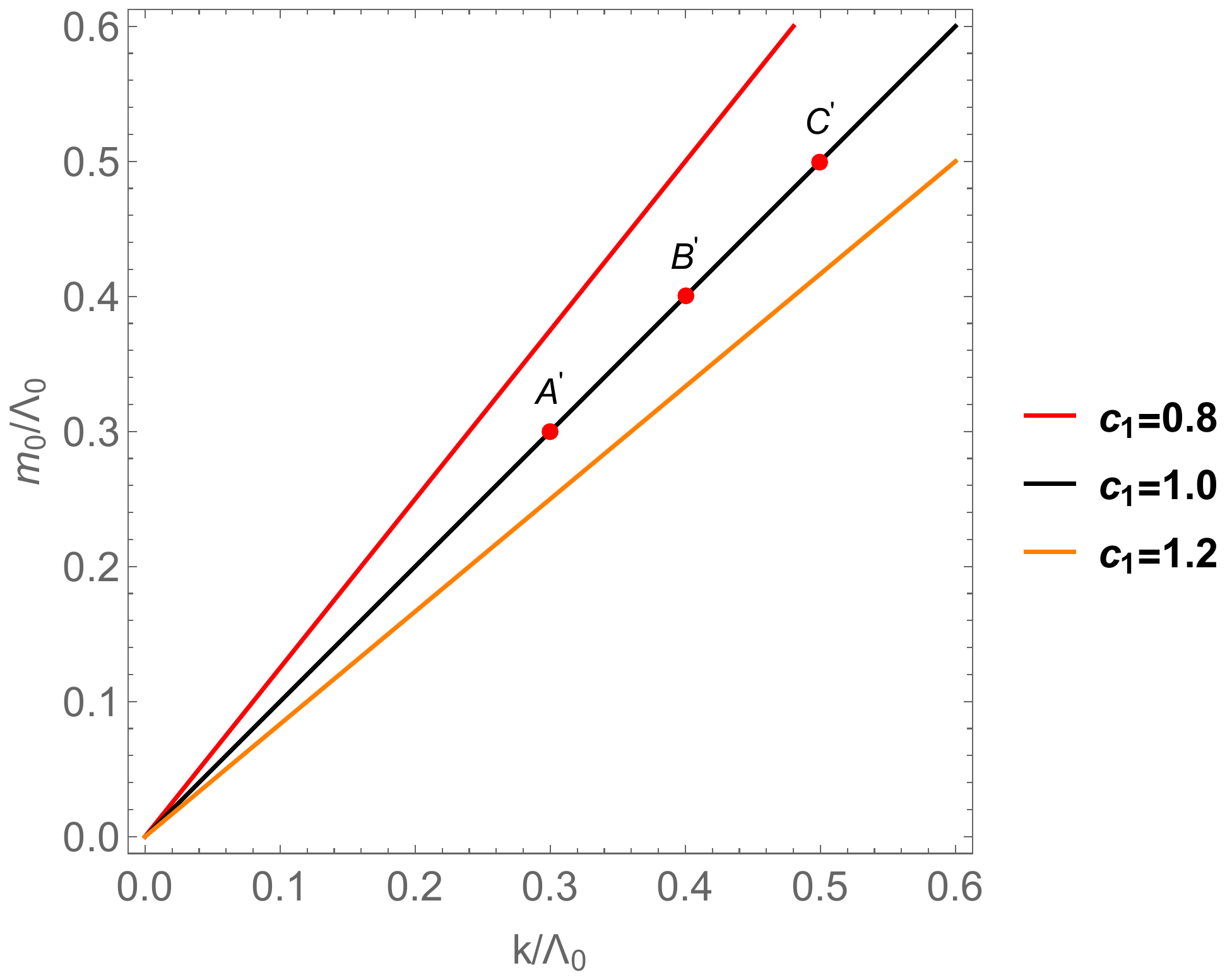}  }

    \caption{{\bf : The numerical parameter space and the physical parameter space:} The figure (a) shows solution lines with $c_1=$0.8(red), 1(black)  and 1.2(Orange) in terms of numerical shooting parameters. The corresponding solution lines in the physical parameter space are shown in figure (b).    
} \label{fig:NsolToPspace}
\end{figure}

Let us clarify how the numerical parameter space is related to the physical parameter space. In order to compare the free energies of those two solutions, we start with a useful quantity defined by: 
\begin{align}
c_1 \equiv \frac{k}{m_0} =  e^{-w_1(\infty)} \frac{\tilde{k}}{\tilde \rho_1}~. 
\end{align}
This parameter is independent of $r_0$ and $r_h$. We show the solution lines with several $c_1$'s in Figure \ref{fig:NsolToPspace}. Each point in Figure \ref{fig:NsolToPspace} (a) corresponds to an AdS Q-Soliton and a Q-lattice black brane. For a given $\Lambda_0$ or $T$, the physical parameters, $m_0$ and $k$, become 
\begin{align}\label{m0k}
&m_0 =\frac{4 \pi   \left(\rho (1)^2-1\right) \tilde{\rho }_1 e^{w_0(\infty )-w_0(1)}}{\rho (1)^2-3}\,\times\left(\Lambda_0~ \text{or}~ T\right)~,~\nonumber\\
&k = \frac{4 \pi \left(\rho (1)^2-1\right) \tilde{k} e^{w_0(\infty )-w_1(\infty )-w_0(1)}}{\rho (1)^2-3} \,\times\left(\Lambda_0~ \text{or}~ T\right)~,
\end{align} 
where we used the  relations (\ref{Temp}), (\ref{ConScale}), (\ref{m0}) and (\ref{kphy}). From this expression, one can see that a numerical solutions can be mapped to physical solutions with different $m_0$ and $k$ by choosing  the value of $\Lambda_0$ or $T$ and, in fact, choosing $r_h$ or $r_0$. See (\ref{Temp}) and (\ref{ConScale}). Taking a fixed $\Lambda_0$ and using numerical solutions, one can show that AdS Q-Solitons corresponding to A, B and C map to A${}'$, B${}'$ and C${}'$ in Figure \ref{fig:NsolToPspace}. For a given $T=\Lambda_0$, so do black branes corresponding to A, B and C.

On the other hand, the black branes corresponding to A, B and C could map to B${}'$ by choosing appropriate temperature $T$'s. In this case, A, B and C have different temperatures. Specifically, the A black brane has  $T=4/3\,\Lambda_0$ and the C black brane has $T=4/5\,\Lambda_0$ while the B black brane has still the same temperature ($T=\Lambda_0$). Thus the left points and the right points of B along the black solid line in Figure \ref{fig:NsolToPspace} (a) describe the higher temperature black branes and the lower temperature black branes with the same $m_0$ and $k$, respectively. Using this mapping, one can compare the Euclidean on-shell action of temperature-varying black branes to that of the AdS Q-soliton for a fixed $\Lambda_0$ at the specific $(k,m_0)$ point in the physical parameter space, Figure \ref{fig:NsolToPspace} (b).

\begin{figure}[ht!]
\centering
    \subfigure[ ]
    {\includegraphics[width=10cm]{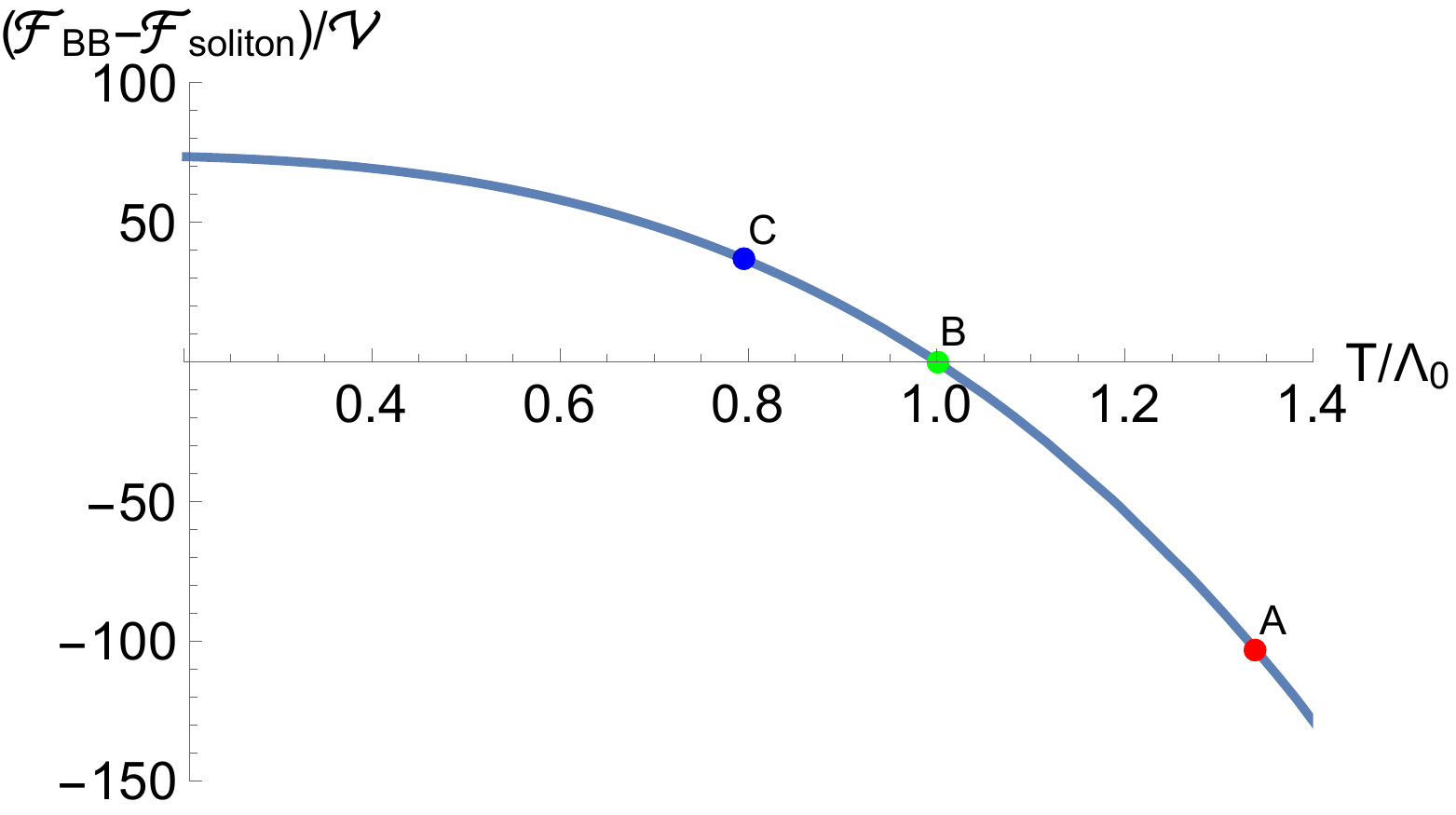}  }

    \caption{{\bf Free Energy Difference at $m_0=k= 0.4\, \Lambda_0$:} The solid line describes the difference of free energies of the temperature-varying black brane along $c_1=1$ line and the AdS Q-soliton given by B in Figure \ref{fig:NsolToPspace} (a). This confirms that the AdS Q-soliton is dominant at $T<\Lambda_0$ while the Q-lattice black brane is preferable at $T>\Lambda_0$.   
} \label{fig:dF}
\end{figure}

By gauge/gravity duality, the free energy is given by $\mathcal{F} = T S_{on-shell}$ in terms of the Euclidean on-shell action (\ref{onShell}). This prescription allows one to find which phase is preferable to the other phase. As a representative case, let us consider $c_1=1.0$ line (The black solid line in Figure \ref{fig:NsolToPspace}). We will focus on the case with $m_0=k=0.4 \,\Lambda_0$. Then the AdS Q-soliton is given by B in Figure \ref{fig:NsolToPspace} (a). Using (\ref{m0k}) and (\ref{onShell}), we obtained the free energy difference between the AdS Q-soltion with $m_0=k=0.4\, \Lambda_0$ and the temperature-varying black branes sharing the same $m_0$ and $k$. The result is presented in Figure \ref{fig:dF}, where one can see that the black brane at the lower temperature ($T <\Lambda_0$) has bigger free energy than the AdS Q-soliton while the AdS Q-soliton does at higher temperature ($T>\Lambda_0$). Also, we have checked that it is always true for the other $c_1$'s.

\begin{figure}[ht!]
\centering
    \subfigure[ ]
    {\includegraphics[width=10cm]{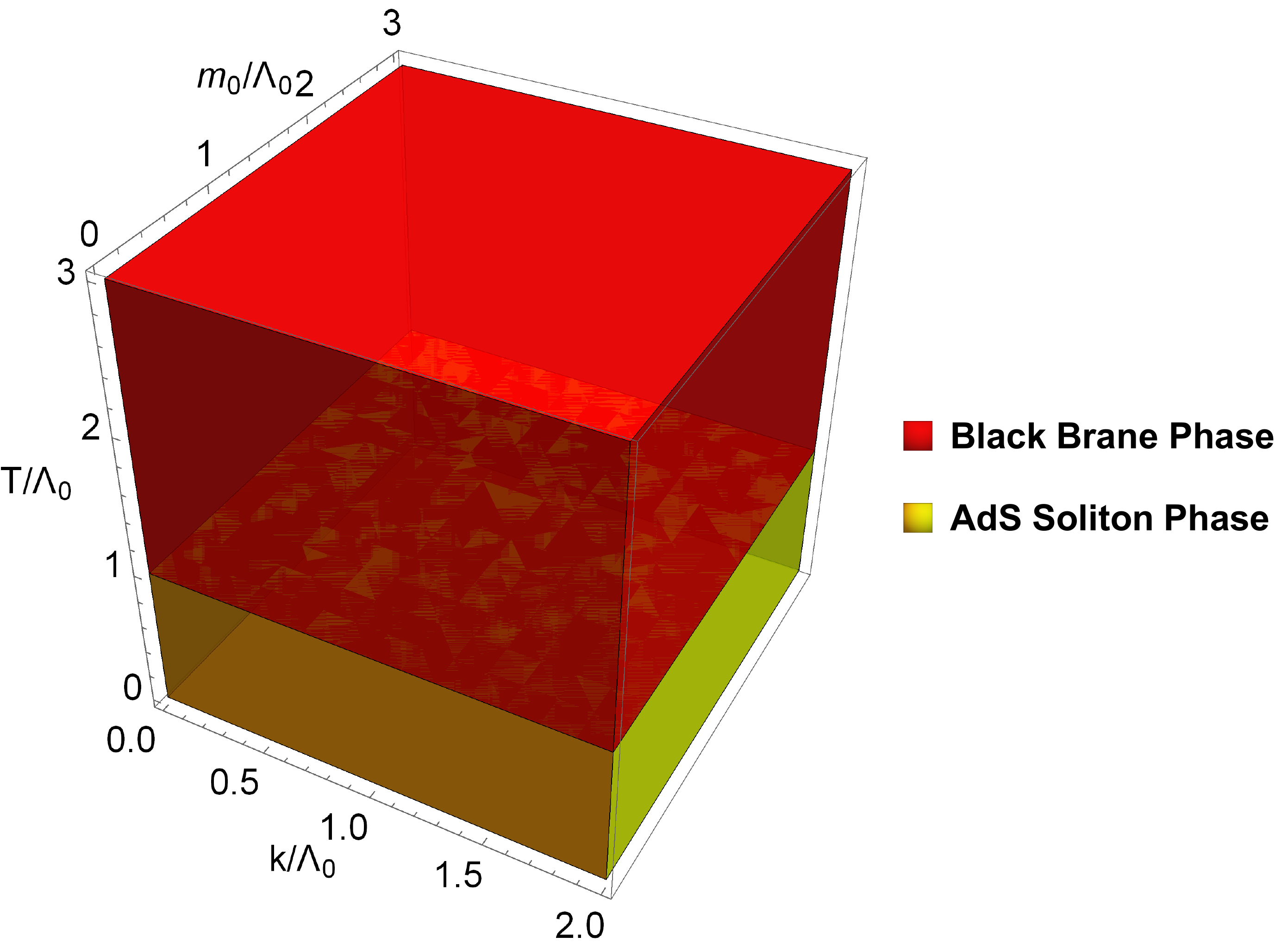}  }

    \caption{{\bf The phase diagram of the gravity model:} The AdS Q-solitons(Confining Phase) have smaller free energies than the Q- lattice black branes(Deconfinement Phase) at the same $m_0/\Lambda_0$ and $k/\Lambda_0$ below $T=\Lambda_0$.    
} \label{fig:phaseD}
\end{figure}

This comparison confirms that AdS Q-solitons are preferable at low temperature, while Q-lattice black branes are dominant at high temperature. Therefore the dual field theory undergoes a confinement-deconfinement phase transition at $T=\Lambda_0$. In the bulk physics, this transition is nothing but the Hawking-Page transition. We provide the phase diagram in Figure \ref{fig:phaseD}.

Now one can ask what physical quantity or order parameter distinguishes these two phases in the dual field theory. One appropriate quantity is the entanglement entropy \cite{Klebanov:2007ws}. In the next section, we study the holographic entanglement entropy(HEE) in the confining phase. The entanglement entropy of the confining phase shows very different behavior from that of the deconfinement phase. See Figure \ref{fig:DeltaS} (a) for an example. As one can see the difference in the figure, a critical length scale appears and characterizes the entanglement entropy of the confining phase. It would be of importance to understand the confining mechanism via this critical length.

As a final discussion of this section, let us see how the gravity solutions are related to the ImABJM models \cite{Kim:2018qle}. In this work we have focused on a special case with the mass function $m(x) = m_0 \sin k x$. At zero temperature, a dual supergravity solution is, so-called, Susy Q solution studied in \cite{Gauntlett:2018vhk}. Such a supersymmetric configuration is obtained by the Q-lattice ansatz fixing the scalar as $z=\mathcal{R}(r)\, e^{i k x}$ and taking only diagonal component of the metric non-vanishing. Our ansatz has been taken to be the similar form of the Susy Q ansatz. See (\ref{BBrane00}) and (\ref{solition01}). Therefore our solutions can be regarded as the finite temperature version of the dual geometry.

\begin{figure}[ht!]
\centering
    \subfigure[ ]
    {\includegraphics[width=5.5cm]{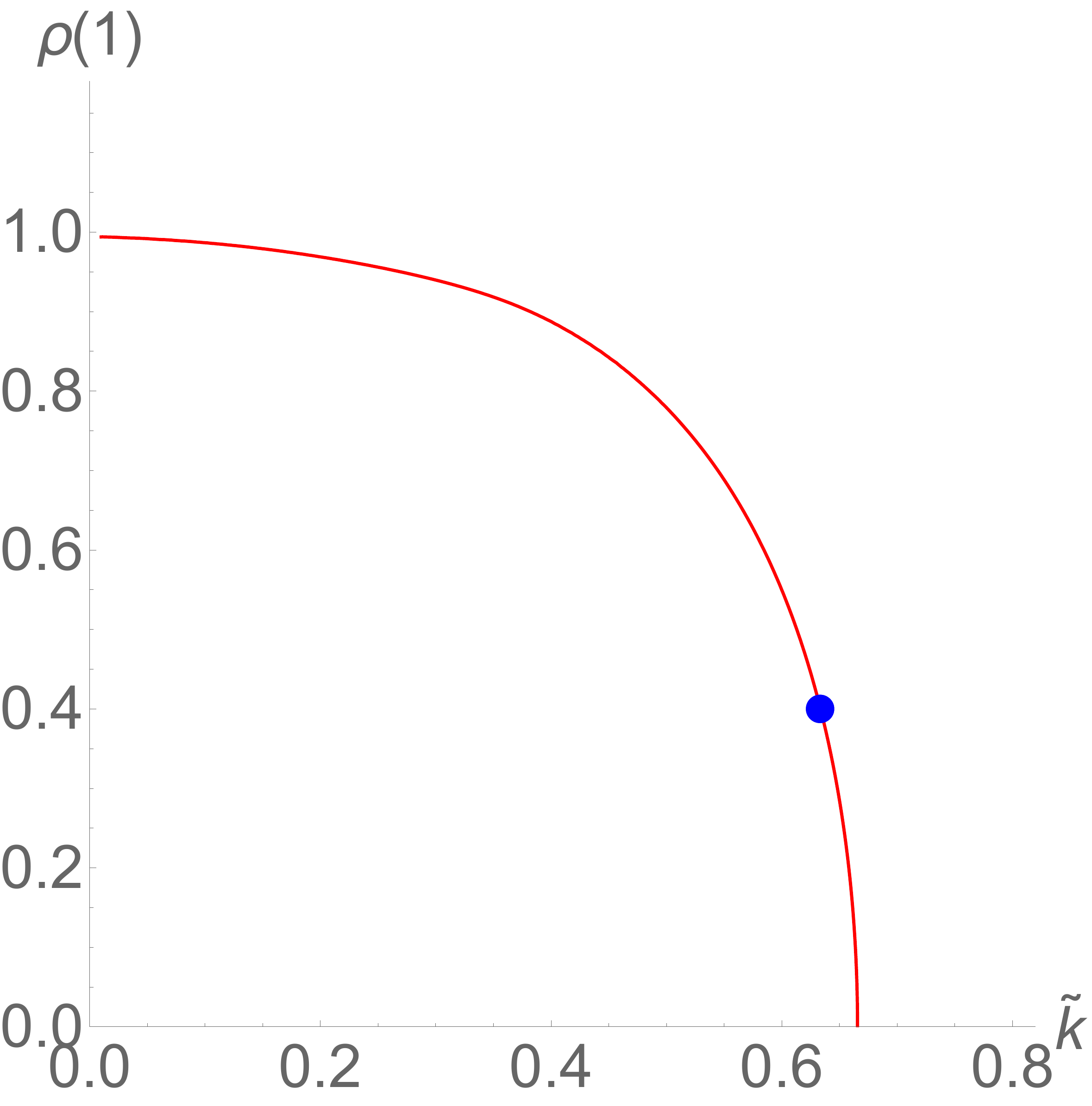}  }
\subfigure[ ]
    {\includegraphics[width=8.5cm]{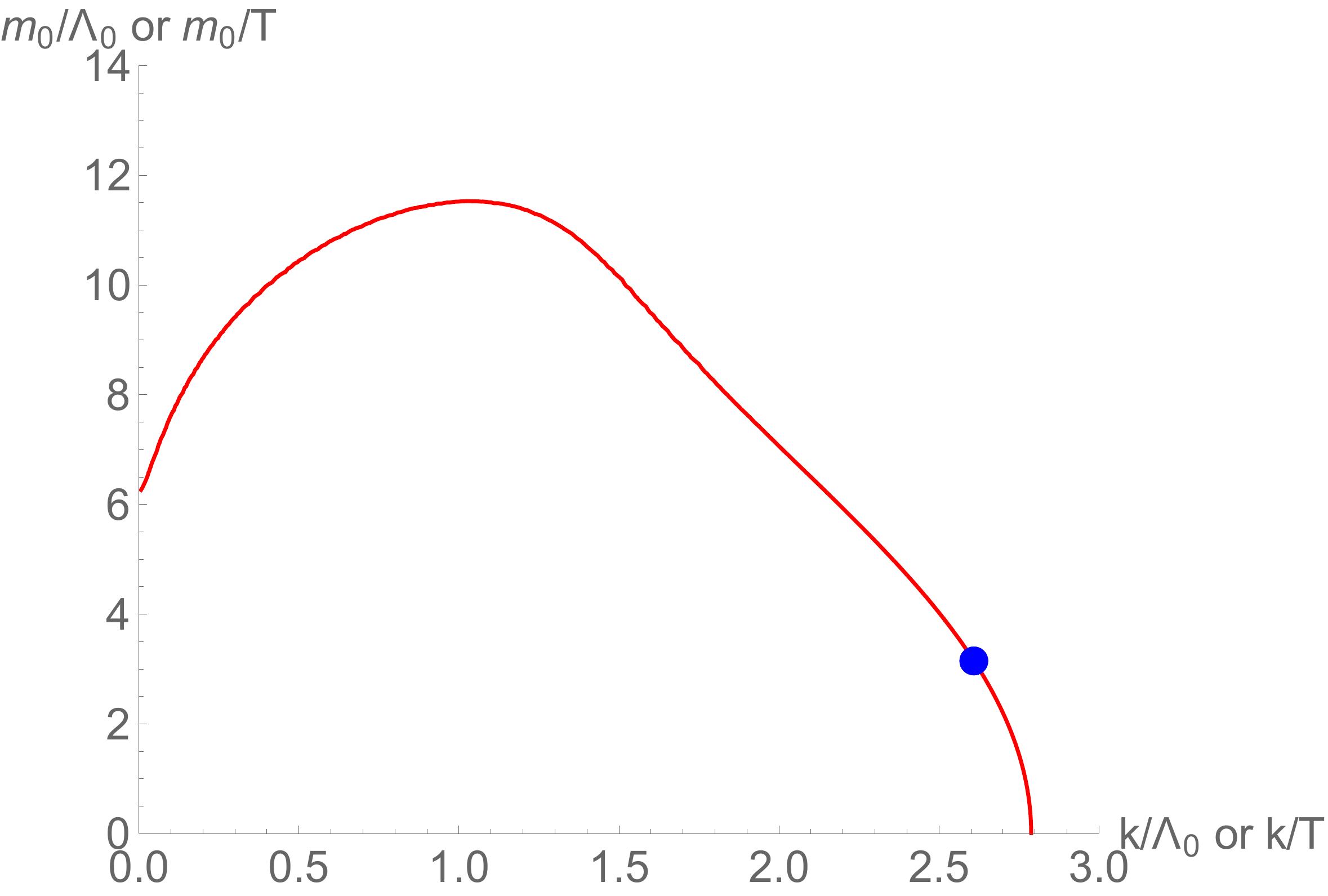}  }

    \caption{{\bf The solution line with $\eta=1$ :} The numerical solution corresponding to the blue dot in the figures is depicted in Figure \ref{fig:HBH01}. Here we took $w_0(1)=w_1(1)= 10^{-4}$.    
} \label{fig:susyQ}
\end{figure}

In the next section, we study modulation effect on the HEE for AdS Q-solitons. As a representative case, we select the $\eta=1$ solution set. In order to pick such solutions, we have to find numerical solutions satisfying 
\begin{align}\label{susyDeformation_cond}
\frac{j_{\mathcal X}}{j_{\mathcal Y}} = k ~~\text{or}~~ \frac{\tilde{\rho}_2}{\tilde{\rho}_1}=-\tilde{k} e^{-w_1(\infty)},
\end{align} 
where the second expression is for the numerical parameter space. We show the numerical solution line satisfying the above condition in Figure \ref{fig:susyQ}. In the next section, we show the characteristics of the HEE based on this solution line. 
 
\section{HEE Probing Confinement}

In the previous section, we have investigated the phase diagram for the dual geometry with a complex scalar hair. Now we explore the confinement phenomena through the entanglement entropy. In order to do this, we rely on the calculation of the HEE~\cite{Ryu:2006bv,Ryu:2006ef} and the discussion on the confinement phenomenon based on \cite{Klebanov:2007ws}. See also \cite{Nishioka:2006gr}. In particular, as we mentioned, we focus on the AdS Q-solitons of the supersymmetrically deformed case depicted in Figure \ref{fig:susyQ}.

\subsection{HEE for the (dis)connected minimal surface}
 We start with the AdS Q-soliton geometry
\eqref{solition01},
\begin{align}
ds^2 =-\frac{r^2}{L^2}  dt^2 +\frac{r^2}{L^2} e^{2\tilde W_1(r)} d{x}^2  +\frac{U(r)}{L^2}e^{2\tilde W_0(r)} d{\chi}^2 + \frac{L^2 dr^2}{U(r)}~,
\end{align}
where we define $\tilde W_i(r) \equiv W_i(r) - W_i(\infty)$ with $i=0,1$.
Under the mapping $X^M = X^M (\sigma^i)$ with coordinates $\sigma^i$'s $(i =1,2)$ on the codimention 2 hypersruface at a constant time slice, 
\begin{align}
x = \sigma_1~, \qquad r = r(\sigma_1)~,\qquad \chi = \sigma_2~.
\end{align}
We obtain the induced metric  $h_{ij} = G_{MN} \frac{\partial X^M\partial X^N}{\partial \sigma^i\partial\sigma^j}$ as
\begin{align}\label{gij}
&h_{11} =  \frac{r^2}{L^2}\, e^{2 \tilde W_1(r)} + \frac{L^2}{U(r)} r'^2~,\qquad h_{22} = \frac{U(r)}{L^2}\, e^{2 \tilde W_0(r)}~,
\end{align}
where $r'\equiv \frac{dr}{d x}$ in this section. 
Then the area of the codimension 2 hypersurface subject to the entangling region of length $l$ in the $x$-direction, while  spanning whole $S^1$ circle of the $\chi$-direction, on the AdS boundary is given by 
\begin{align}\label{gammaA}
\gamma_A &= \int d^2\sigma\sqrt{\det h_{ij}}
= 2\chi_p\int_0^{\frac{l}{2}} d  x \sqrt{H(r)}\,\sqrt{1+ \beta (r)r'^2}~,  
\end{align}
where $\chi_p\equiv \frac{4\pi L^2}{U'(r_0)} e^{-\tilde W_0(r_0)}$ denotes the periodicity of the coordinate $\chi$ of the $S^1$ circle and 
\begin{align}\label{Hrbetar}
H(r)= \frac{r^2 U(r)}{L^4}\, e^{2(\tilde W_0(r) + \tilde W_1(r))}~,\qquad
\beta(r) = \frac{L^4}{r^2 U(r)}\, e^{-2 \tilde W_1(r)} = \frac{e^{2\tilde W_0(r)}}{H(r)}~.  
\end{align}
Boundary conditions for the hypersurface at the AdS$_4$ boundary are imposed by
\begin{align}\label{xhatbnd}
r( x)|_{ x = -\frac{l}{2}}=r( x)|_{ x 
= \frac{l}{2}}= \infty~.
\end{align}

Applying the RT conjecture~\cite{Ryu:2006bv,Ryu:2006ef} to \eqref{gammaA}, we obtain the HEE as
\begin{align}\label{S_A}
S_A \equiv \frac{{\rm Min}(\gamma_A)}{4 G} 
=\frac{\chi_P}{2 G}\int_{0}^{\frac{l}{2}} dx\, {\cal L}_A~, 
\end{align}
where $G$ is the four-dimensional gravitational constant and the Lagrangian ${\cal L}_A$ is defined as
\begin{align}\label{calLA}
{\cal L}_A \equiv  \sqrt{H(r)}\, \sqrt{1+\beta(r)r'^2}~.
\end{align}
The Hamiltonian density of \eqref{calLA}, which is a constant, is given by
\begin{align}\label{HamA}
{\cal H}_A = - \frac{ H(r) }{ \sqrt{1+\beta(r)r'^2}} = -  \sqrt{H_*}= {\rm constant}~,
\end{align}
where $r_* \equiv r(0)$ denotes the location of the tip of the minimal surface and $H_*\equiv H(r_*)$ using the fact $r'(0) = 0$.
From \eqref{HamA}, we obtain the equation of motion for $r(x)$ as, 
\begin{align}\label{rpxhat}
r'(r) = \frac1{\sqrt{\beta(r)}}\,
\sqrt{\frac{H(r)}{H_*}-1}\,~. 
\end{align}
Then the entangling length $l$ is given by 
\begin{align}\label{lrstar}
l(r_*) = 2\sqrt{H_*}\int_{r_*}^\infty dr \frac{\sqrt{\beta (r)}}{\sqrt{H(r)- H_*}}~.
\end{align}
Here we note that the relation \eqref{lrstar}, which was obtained from \eqref{rpxhat}, is satisfied for the range $r_* >r_0$.\footnote{ Naturally, $r'(r)$ over the range $r<r_0$ in \eqref{rpxhat} is always divergent at $r_* = r_0$. } 
Therefore, for the $r_* > r_0$ case, the equation \eqref{rpxhat} is well-defined over the whole region of $r\ge r_*$, and so the solution describes the {\it connected} minimal surface. Plugging \eqref{rpxhat} into \eqref{S_A} for $r_* >r_0$, we obtain the HEE for the {\it connected} minimal surface,
\begin{align}\label{SAcon}
S_A^{({\rm con})} = \frac{\chi_p}{2G_N}\int_{r_*}^{r_\infty} dr \frac{H(r)\sqrt{\beta (r)}}{\sqrt{H(r)- H_*}}~,
\end{align}
where we introduce $r_\infty$ representing the UV cutoff in the asymptotic limit.
On the other hand, for the case $r_*=r_0$, $r'(r)\to \infty$ with $r>r_*$ and $r'(r)_{r=r_*}$ is not well-defined. This indicates that the corresponding minimal surface is {\it disconnected}. The HEE for the {\it disconnected} one becomes 
\begin{align}\label{SAdiscon}
S_A^{({\rm discon})} = \frac{\chi_p}{2G_N}\int_{r_0}^{r_\infty} dr \sqrt{\beta(r) H(r)}~.
\end{align}

\subsection{Confinement for supersymmetric deformation}

\begin{figure}[ht!]
\centering
    \subfigure[ ]
    {\includegraphics[width=8.0cm]{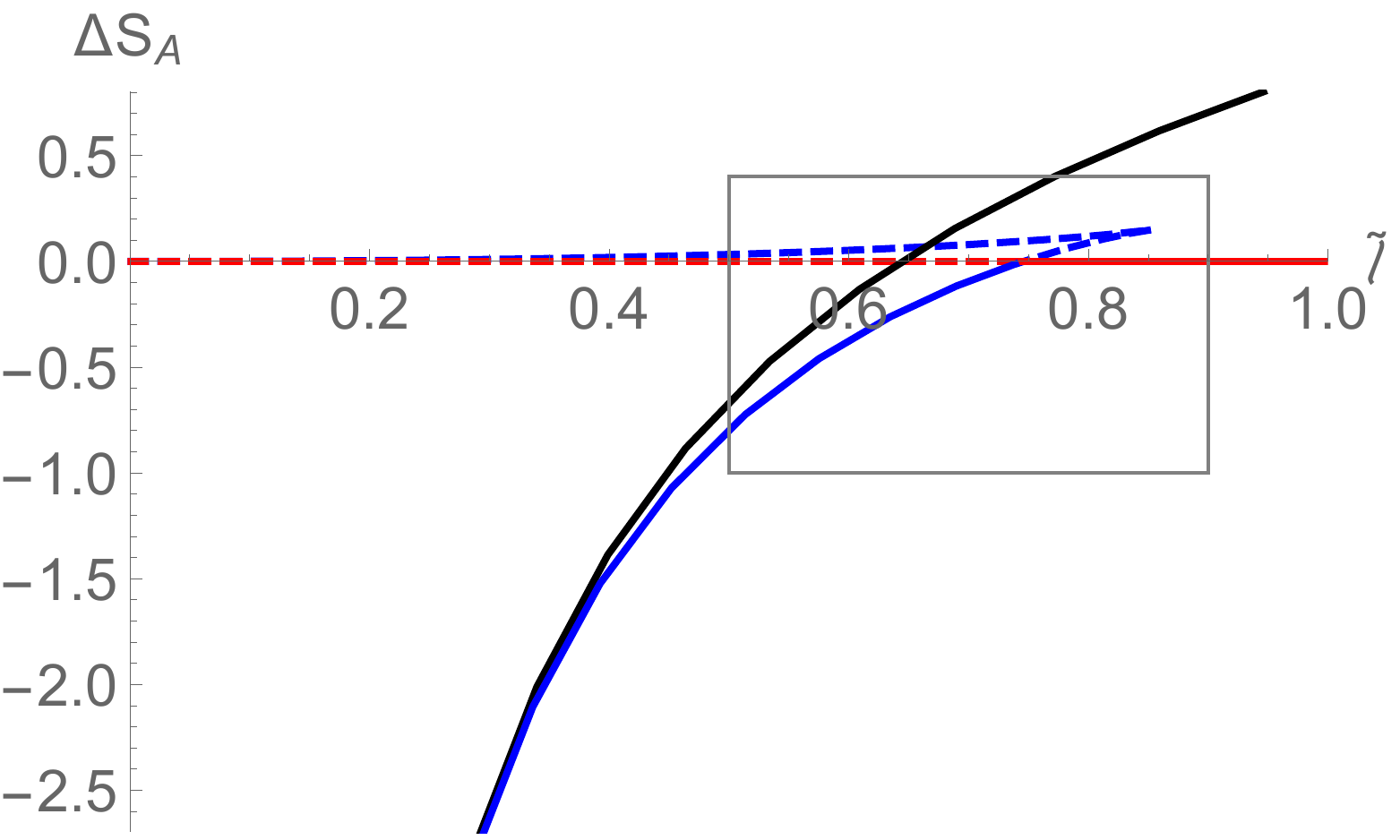}  }
    \subfigure[ ]
    {\includegraphics[width=8.0cm]{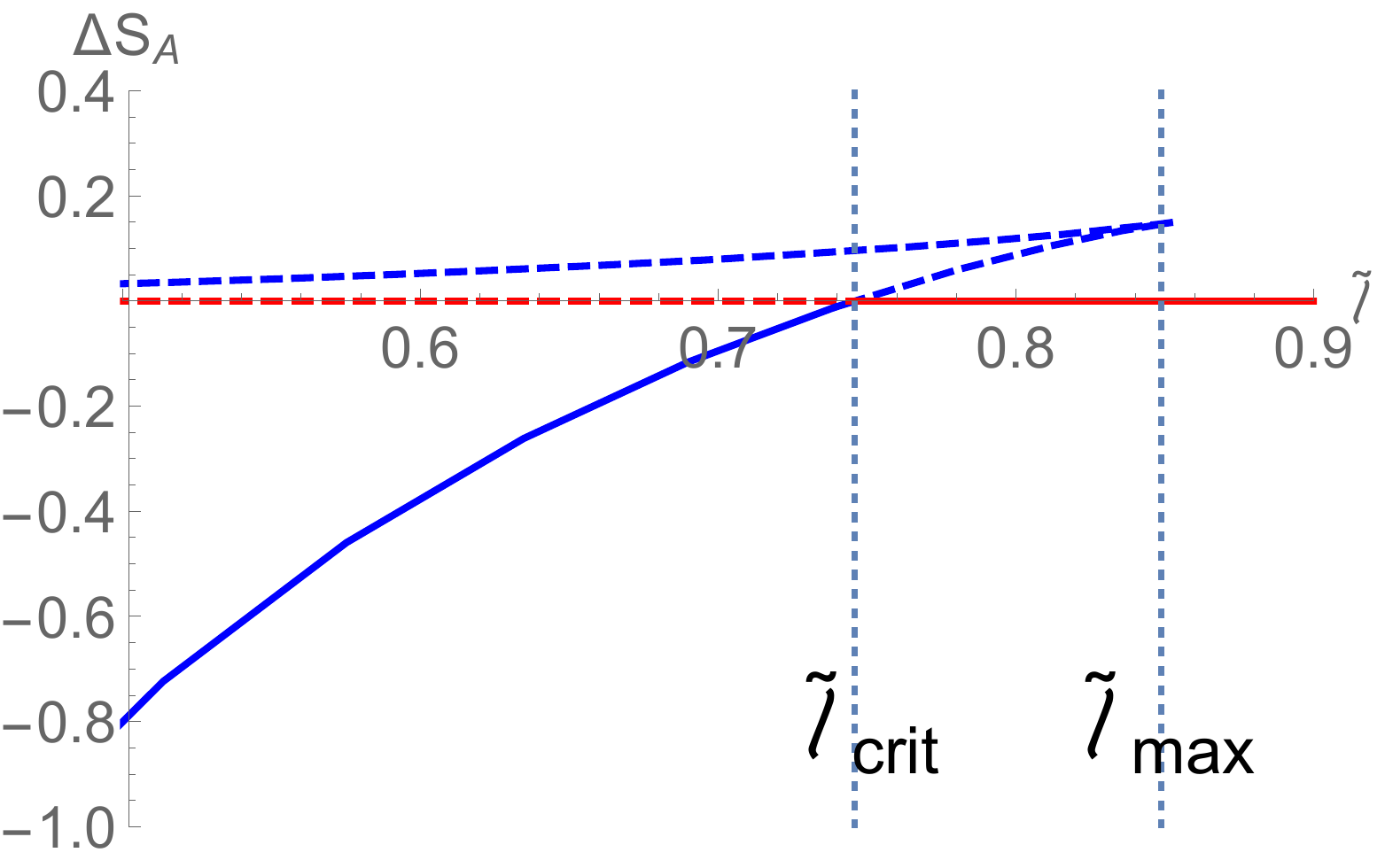}  }

    \caption{ {\bf HEEs for an AdS Q-Soliton and a Q-lattice black brane:}  These figures depict HEE curves in the background geometry corresponding to the blue dot in Figure \ref{fig:susyQ}. The black solid line in (a) shows the entanglement entropy curve for the Q-lattice black brane, where the values were subtracted by $S_A^{\rm (discon)}$. The blue and red solid lines show the entanglement entropy for the AdS Q-soliton. One can notice that there is a drastic change at $\tilde{l}=\tilde{l}_{\rm crit}$. The figure (b) is the enlarged view of the rectangle in (a). The blue lines show the HEE given by the connected surface and the red lines are nothing but $S_A^{\rm (discon)}$ of the disconnected surface. } \label{fig:DeltaS}
\end{figure}

As we will confirm in this subsection, the AdS Q-solitons show a certain confining behavior studied in \cite{Klebanov:2007ws}. This behavior can be shown by evaluating the HEE. The HEE of the AdS Q-solitons in terms of the entangling length $l$ shows quite different curves from those of the Q-lattice black branes. Also the curves are not smooth and each curve has a scale $l_{\rm crit}$ showing a border between connected and disconnected minimal surfaces. A result with a specific $l_{\rm crit}$ is plotted in Figure \ref{fig:DeltaS}. The dimensionless length $\tilde{l}$ in the figures is defined in (\ref{scal}). This subsection is devoted to explaining how one can obtain the HEE curves for the AdS Q-solitons.

\begin{figure}[ht!]
\centering
    \subfigure[ ]
    {\includegraphics[width=7.5cm]{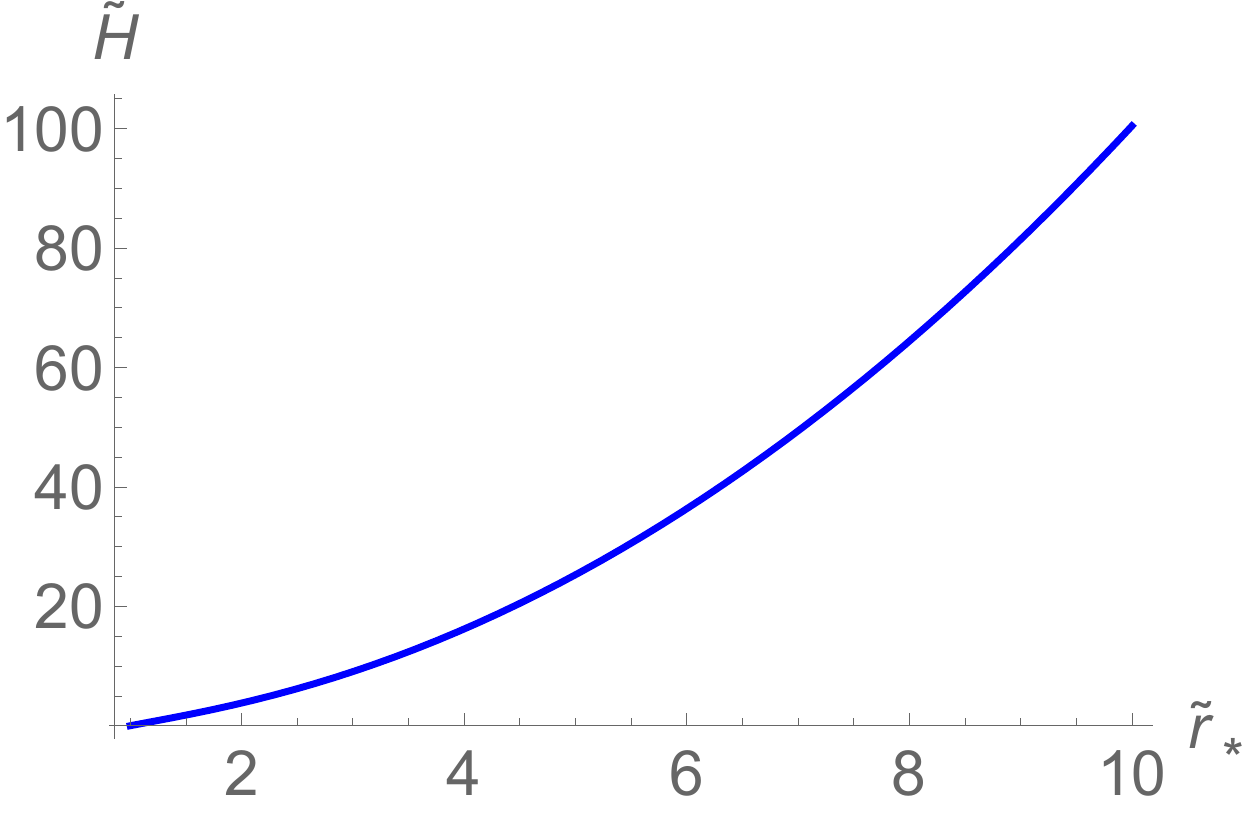}  }
\subfigure[ ]
{\includegraphics[width=7.5cm]{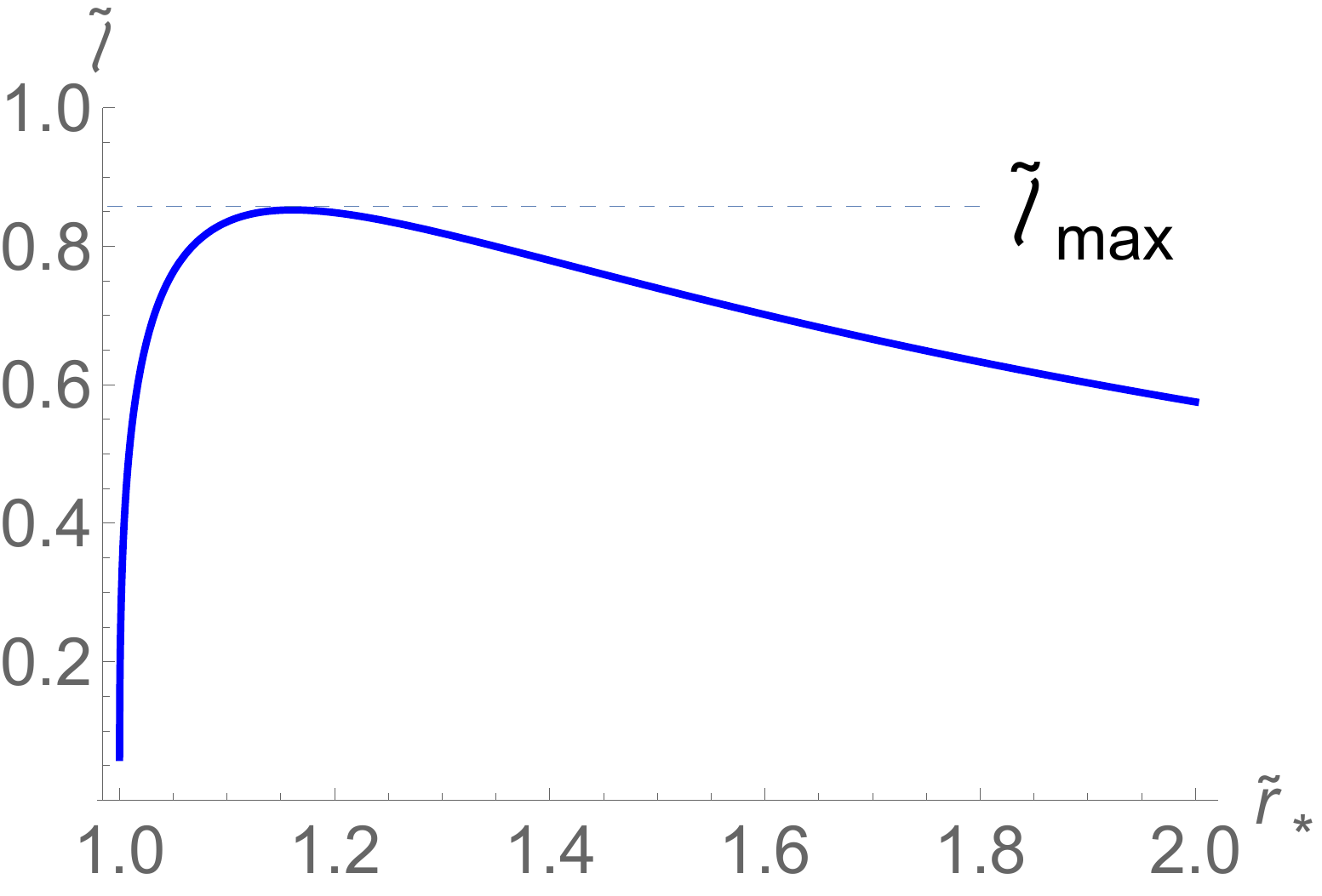}  }

    \caption{{\bf $\tilde{H}(\tilde{r}_*)$ and $\tilde{l}(\tilde{r}_*)$ for a connected surface :} Here the dimensionless quantities are defined in (\ref{scal}) and (\ref{scal2}).  } \label{fig:Hltilde}
\end{figure}

The characteristic behavior of the HEE in the AdS Q-soliton geometry (\ref{solition01}) heavily depends on $H(r)$ in \eqref{lrstar}. As one can see through the definition of $H(r)$ given in \eqref{Hrbetar}, the behavior of the function $H(r)$ is governed by the metric function $U(r)$ in \eqref{solition01}, since the metric functions $e^{2\tilde W_i(r)}$ $(i=0,1)$ are finite and do not show any drastic change over the range $r_0\le r$. 
The $U(r)$ starts from zero at the tip ($r=r_0$), monotonically increases as $r$ increases, and diverges as $U(r)\to r^2$ in the asymptotic region. The function $H(r)$ behaves in the same manner. 
See Figure \ref{fig:Hltilde}.
In the $r_*\to r_0$ limit, $H_*$ approaches zero, and then, $l(r_*)_{r_*\to r_0} \approx 0$ due to the overall factor $\sqrt{H_*}$ in \eqref{lrstar}. On the other hand, 
in the opposite limit $r_*\to \infty$, we have also $l(r_*)_{r_*\to \infty}\approx 0$ due to the short integration range in the right-hand side of \eqref{lrstar}. 
In the intermediate region $r_0< r_* < \infty$, however, $l(r_*)$ should have some finite positive value.  
For this reason, one can argue that there should be a maximum value of $l$ denoted as $l_{{\rm max}}$. See Figure \ref{fig:Hltilde}. The existence of $l_{{\rm max}}$ implies that a regular solution satisfying \eqref{rpxhat}, which is known as the {\it connected} solution, cannot cover the case with $l> l_{{\rm max}}$. In this latter case, only singular solution is possible, which is known as the {\it disconnected} solution having a singular behavior with $r'|_{l>l_{{\rm max}}}= \infty$. As we argued earlier, this {\it disconnected} one is obtained by choosing $r_* = r_0$ in \eqref{rpxhat}\footnote{In fact, the minimal surface for the {\it disconnected} case is given by $x={\rm const}$. So the corresponding surface is $x=\pm l/2$ which clarifies the adjective ``disconnected''.}.

This means that there are two values of HEE for the {\it connected} and {\it disconnected} ones, satisfying the boundary condition \eqref{xhatbnd}. Comparing these two values, one has to choose the smaller value of the HEEs for a given $l$. By increasing the size $l$, it seems that there exists a critical value of $l=l_{{\rm crit}}$, where a kind of first order phase transition occurs between the {\it connected} and {\it disconnected} surfaces. This is the signal for the confinement-deconfinement  phase transition in the dual quantum field theory~\cite{Klebanov:2007ws}.

We confirm this characteristic behavior for the AdS Q-soliton geometry \eqref{solition01} by computing HEE numerically.
To do that, we rescale coordinates and functions in terms of dimensionful parameters and define dimensionless quantities for numerical works. 
As we did in previous sections, we rescale those as
\begin{align}\label{scal}
&r = r_0\tilde r~, \quad y = \frac{L^2}{r_0} \tilde y~,  \quad l(r_*) = \frac{L^2}{r_0} \tilde l (\tilde r_*)~,\quad 
U(r) = r_0^2 u(\tilde r)~,
\nn \\
& W_0(r) = w_0(\tilde r)~, \quad W_1(r) = w_1(\tilde r)~,\quad H(r) = \frac{r_0^4}{L^4}\tilde H(\tilde r)~,\quad 
\beta(r) = \frac{L^4}{r_0^4}\tilde \beta (\tilde r)~,
\end{align}
where
\begin{align}\label{scal2}
\tilde l(\tilde r_*) &= 2  \sqrt{\tilde H_*}
\int_{\tilde r_*}^\infty d\tilde r \frac{\sqrt{\tilde \beta (\tilde r)}}{  \sqrt{\tilde H(\tilde r) -\tilde H_*}}~,
\nn \\
\tilde H(\tilde r) &= \tilde r^2 u(\tilde r) e^{2(\tilde w_0(\tilde r) + \tilde w_1(\tilde r))}~,
\nn \\
\tilde \beta(\tilde r) &= \frac{e^{2\tilde w_0(\tilde r)}}{\tilde H(\tilde r)}
\end{align}
with $\tilde w_i(\tilde r) = w_i(\tilde r) - w_i(\infty)$.

Now, we are ready to find the difference of HEEs for the {\it connected} and {\it disconnected} solutions. Since the divergent behaviors of $S_A^{{\rm (con)}}$ and $S_A^{{\rm (discon)}}$ are the same in the $\tilde r_\infty\to \infty$ limit, their difference $\Delta S_A$ is well-defined and finite. 
Plugging the relations into \eqref{SAcon} and \eqref{SAdiscon}, we obtain 
\begin{align}\label{DelSA}
\Delta S_A \equiv S_A^{{\rm (con)}}- S_A^{{\rm (discon)}}= \frac{\tilde \chi_p}{2 G_N}\left[\int_{\tilde r_*}^{\tilde r_\infty} d\tilde r e^{\tilde w_0(\tilde r)} \left(\frac{1}{\sqrt{1-\frac{\tilde H_*}{\tilde H(\tilde r)}}}-1\right) -\int_1^{\tilde r_*} d\tilde r e^{\tilde w_0(\tilde r)}\right]~,
\end{align}
where $\tilde\chi_p = \frac{4\pi L^2}{u'(1)}\, e^{\tilde w_0(1)}$ with $u'(1) = \frac{U'(r_0)}{r_0}$. 
As one can see in Figure \ref{fig:DeltaS}, $\Delta S_A(\tilde l)<0$ for $\tilde l <\tilde l_{{\rm crit}}$, i.e., $S_A^{{\rm (con)}}$ is smaller than $S_A^{{\rm (discon)}}$. 
Therefore, the HEE in this range of $\tilde l$ becomes $S_A^{{\rm (con)}}$. On the other hand, in the range $\tilde l> \tilde l_{{\rm crit}}$, the HEE becomes $S_A^{{\rm (discon)}}$. The solid line of Figure \ref{fig:DeltaS} shows an entanglement entropy curve  for the confining phase. As we discussed previously, a drastic change of HEE looking like a first order phase transition occurs under varying the entanglement length $l$ near $l_{\rm crit}$. The numerical work for the AdS Q-soliton \eqref{solition01} indicates that the phase transition near $l_{{\rm crit}}$ is qualitatively similar to the one for the confinement phenomenon discussed in \cite{Klebanov:2007ws}. As we mentioned before, one can notice that this feature doesn't appear in the black branes. See the black solid line for the Q-lattice black brane corresponding to the blue dot in the Figure \ref{fig:DeltaS} (a).

\subsection{Characteristic length $l_{\text{crit}}$ in the parameter space of $m_0$ and $k$}

So far, via the HEE study in the AdS Q-solitons, we have shown that the entanglement entropy curve seems to indicate a first order phase transition related to the confinement-deconfinement phase transition. It would be interesting to investigate such a phenomenon in the backgrounds corresponding to the ${\cal N} = 3$ ImABJM model with the mass function $m(x) = m_0 \sin  kx$. To do that, we pick up solutions of the supersymmetric deformation by choosing $\eta=1$. This set of solutions was already obtained in the Figure \ref{fig:susyQ}. By using these solutions, we would like to expose how the confinement phenomenon can be affected by the modulation effect.

\begin{figure}[ht!]
\centering
    \subfigure[ ]
    {\includegraphics[width=8.0cm]{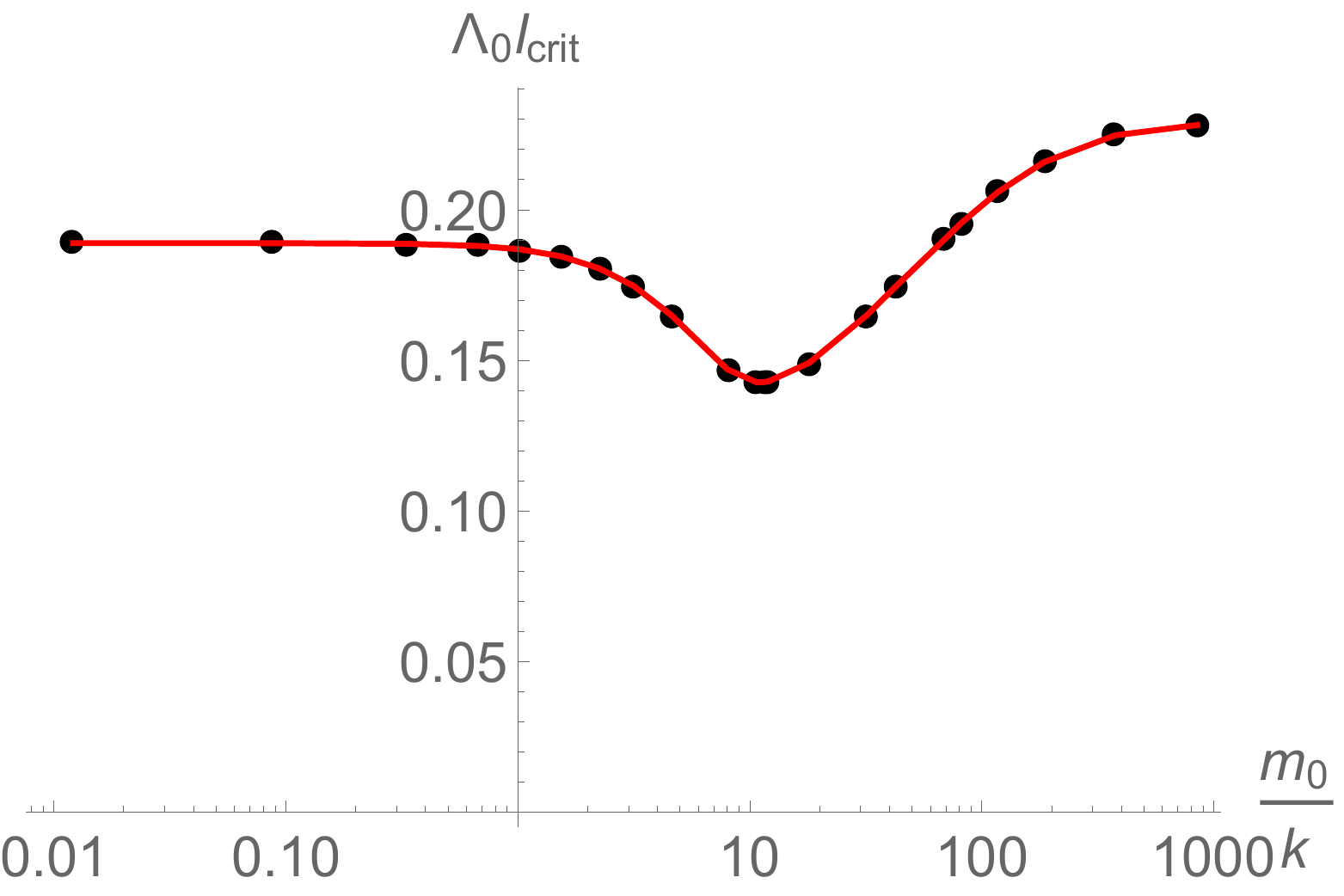}  }

    \caption{{\bf Characteristic length for AdS Q-solitons:} This result is based on the set of solutions (The red solid line in Figure \ref{fig:susyQ}) corresponding to the supersymmetric deformation.} \label{fig:lcritical}
\end{figure}

In order to show the modulation effect, we consider the critical length $l_{\rm crit}$ showing the strength of confinement. The critical length decreases as the period $\chi_p$ of the AdS soliton decreases. In this limit the AdS soliton becomes more and more different from the black brane or the pure AdS space. Thus it is legitimate to argue that the small critical length indicates enhancement of the confinement and vice versa. We find the behavior of $l_{\rm crit} \Lambda_0$ to fix the confining scale and present the result in terms of another dimensionless parameter $\frac{m_0}{k}$, which represents the ratio between the amplitude and modulation of the inhomogeneity in the mass function.
The result using the data of the solution line in Figure \ref{fig:susyQ} is given in Figure \ref{fig:lcritical}. Interestingly, there exists the minimum characteristic length of the confinement at a specific ratio of $m_0$ and $k$, though the characteristic length approaches constant values in the limits of $m_0\ll k$ and $m_0\gg k$. Both limits actually correspond to the pure AdS soliton and the hairy AdS soltion without modulation for finite $k$, respectively. Therefore the existence of minimum $l_{\rm crit}$ in the intermediate region tells us that the modulation effect enhances the confinement. It deserves to investigate the nature of this phenomenon in more detail, which is beyond the scope of this paper.

\section{Discussion}

In this paper, we started with the action \eqref{bulk_action} and  employed the Q-lattice ansatz to obtain new background solutions. The black branes with a complex scalar hair and the AdS Q-solitons were obtained as the background solutions. These geometries could be dual to a deformed ABJM theory by $\int d^3 x \left(\, m'(x) \mathcal{O}_{\mathcal X} + m(x) \mathcal{O}_{\mathcal Y} +\ldots \,\right)$ in the finite temperature, where the dotted line denotes operator deformations which cannot be protected in the large $N$ limit. We found the regular solutions using a numerical method and identified the various physical quantities in terms of numerical parameters. The Q-lattice black branes describe the deconfinement phase of the dual field theory, while the AdS Q-solitons are regarded as gravity duals of the confining phase. These two phases constitute the phase diagram by comparing the free energies of both geometries. It was shown that the AdS Q-soliton is preferable in the low temperature and the Q-lattice black branes dominate in the high temperature region. It would be interesting to study the thermodynamics of the Q-lattice black branes. In fact, the corresponding black brane thermodynamics is rich and contains nontirival physical implication. We leave it as our next work and it will be reported soon\cite{Hyun:2019juj}.  

The numerical solutions of various background geometries allow various $\eta$'s. When $\eta=1$, the solution is clearly the gravity dual to the ${\cal N} = 3$ ImABJM model with the mass function given by $m(x) =m_0 \sin k x$. Thus our work is the first study on this ImABJM model in the finite temperature regime. 

We have studied the HEE in our background solutions. The figure \ref{fig:DeltaS} shows the entanglement entropy curves for certain Q-lattice black brane and AdS Q-soliton as a representative case. The behaviors of those two curves are quite different. Also the AdS Q-soliton has a characteristic length scale which may be important to figure out the confining phase of the ImABJM theory. We have also studied the modulation effect on the characteristic length $l_{\text{crit}}$. We conclude that the modulation enhances the confinement. In addition, we notice that there exists a special value of the critical length $l_{\rm crit}$ in the limit $m_0\gg k$ in Figure \ref{fig:lcritical}. This phenomenon may indicate a possibility that the ${\cal N} = 6$ mABJM theory with a {\it constant} mass parameter has the confinement-deconfinement phase transition at a certain low energy scale. It is desirable to clarify this possibility as a future direction.

It would be very interesting  to find solutions beyond the Q-lattice ansatz, though expected to be difficult in the numerical analysis. We leave the problem as a future work. In addition higher dimensional supergravity models are  interesting in applying our approach and it also deserves to try the different types of the entangling regions such as a circular entangling surface. The physical implication of the characteristic length in the dual field theory would be clarified in the near future.

\section*{Acknowledgments}

This work was supported by the National Research Foundation of Korea(NRF) grant with grant number NRF-2016R1D1A1A09917598 (B.A., S.H., S.P.), NRF-2019R1A2C1007396 (K.K., B.A.), and    NRF-2017R1D1A1A09000951, NRF-2019R1F1A1059220, NRF-2019R1A6A1A10073079 (O.K.). K.K. acknowledges the hospitality at APCTP where part of this work was done.

\section*{Appdendix}

\subsection*{A. Holographic Renormalization}

In \cite{Gauntlett:2018vhk}, the authors introduced the finite actions as follows:
\begin{align}
\mathcal{S}_{tot} = \mathcal{S}_B + \frac{1}{8\pi G} \int_{\partial \mathcal M} d^3 {x} \sqrt{-\gamma}\,K + \mathcal{S}_{ct} + \mathcal{S}_L~,
\end{align}
where $\mathcal{S}_B$ is given in (\ref{bulk_action}) and $K$ is the Gibbons-Hawking therm, which is defined by $K= \gamma^{\mu\nu}K_{\mu\nu}$ and $K_{\mu\nu}=\frac{1}{2N}\gamma'_{\mu\nu}$. The induced metric $\gamma_{\mu\nu}$ is based on the following ADM decomposition:
\begin{align}
ds^2 = N(r)^2 dr^2 + \gamma_{\mu\nu} d {x}^\mu d {x}^\nu~,
\end{align}
where the boundary coordinates ${x}^\mu$'s are normalized to have $\eta_{\mu\nu}$ as the boundary metric. $S_{ct}$ and $S_L$ are given as follows
\begin{align}
&\mathcal{S}_B =\frac{1}{16\pi G}\int dr d^3  x \sqrt{-g}\,\mathcal{L}_{bulk}~,\label{bulk_action_2}\\
&\mathcal{S}_{ct} =-\frac{4}{16\pi G L} \int d^3  {x}\sqrt{-\gamma} \left(1 +\frac{1}{2}|z|^2  \right)~,~\\
&\mathcal{S}_L =\frac{4}{16\pi G L } \int d^3  {x} \sqrt{-\gamma}  \left( r \mathcal{X}\partial_r \mathcal{X} + \mathcal{X}^2 \right)~,
\end{align}
Here the $\mathcal{X}$ is the real part of the complex scalar field, which is dual to the dimension one operator. The complex field $z$ is given by
\begin{align}
z = \mathcal{X}+i\,\mathcal{Y} = \mathcal{R}(r) \cos  {k}  {x}  + i \,\mathcal{R}(r) \sin  {k}  {x}~.
\end{align}
More explicitly we have the relations, 
\begin{align}\label{Counter00}
\mathcal{S}_{ct} + \mathcal{S}_L = \frac{1}{16\pi G L} \int d^3  {x} \sqrt{-\gamma}\left( -4 -2 \mathcal{Y}^2 -2 \mathcal{X}^2 + 4 r \mathcal{X}\partial_r \mathcal{X} +4 \mathcal{X}^2  \right)~.
\end{align}
The VEVs of the operators for black branes are given by
\begin{align}
\left<\mathcal{O}_\mathcal{X}\right>& =\lim_{r\to \infty} \frac{1}{16\pi G } \frac{L^4}{r^2} \left(- \sqrt{-g}\frac{4}{1-|z|^2} \nabla^r \mathcal{X} + \frac{4}{L}\sqrt{-\gamma}\left( r \partial_r \mathcal{X} + \mathcal{X} \right) \right)~
\nonumber\\
&=\frac{4\tilde\rho_1\,}{16\pi G}\, r_h \cos  {k}  {x}~,
\label{vevX}\\
\left<\mathcal{O}_\mathcal{Y}\right> &=\lim_{r\to \infty}\frac{1}{16\pi G } \frac{L^2}{r} \left(- \sqrt{-g}\frac{4}{1-|z|^2} \nabla^r \mathcal{Y} - \frac{4}{L}\sqrt{-\gamma}\mathcal{Y} \right)\nonumber\\
&= \frac{4\tilde\rho_2  }{16\pi G L^2}\,  r_h^2  \sin  {k} {x}~.\label{vevY}
\end{align}
Replacing $r_h$ with $r_0$, the above expressions in \eqref{vevX} and \eqref{vevY} are still valid for AdS solitons. In addition, the boundary energy momentum tensor can be evaluated as
\begin{align}
\left<\mathcal{T}^{\mu\nu}\right>=\lim_{r\to \infty} \frac{2}{16\pi G} \frac{r^5}{L^5} \left(-K^{\mu\nu} + K \gamma^{\mu\nu} + \frac{1}{2L} \gamma^{\mu\nu}\left( -4 -2|z|^2 +4r \mathcal{X} \partial_r \mathcal{X} +4\mathcal{X}^2 \right) \right)\,.
\end{align}
Then it turns out that components of the boundary energy momentum tensor for black branes are given by
\begin{align}
& \left< \mathcal{T}^{  t  t} \right> = \frac{r_h^3}{16\pi G L^4} (2 \tilde{m} + 6 \tilde{w}_{1,3} + 8 \tilde{\rho}_1 \tilde{\rho}_2 - 4 \tilde{\rho}_1 \tilde{\rho}_2 \sin^2{ {k} {x}})~,\label{HenergyBH}\\
&\left< \mathcal{T}^{  x  x}\right> =  \frac{r_h^3}{16\pi G L^4} (\tilde{\mathit{m}} +6 \tilde{w}_{1,3}+4 \tilde{\rho }_1 \tilde{\rho }_2  \sin^2{ {k} {x}})~,  \\
&\left< \mathcal{T}^{  y  y}\right> =  \frac{r_h^3}{16\pi G L^4} (\tilde{\mathit{m}} +4 \tilde{\rho }_1 \tilde{\rho }_2  \sin^2{ {k} {x}})~.
\end{align}
On the other hand, the AdS soliton background has dual energy-momentum tensor as follows:
\begin{align}
&\left< \mathcal{T}^{  t  t}\right> =  \frac{r_0^3}{16\pi G L^4} (-\tilde{\mathit{m}} -4 \tilde{\rho }_1 \tilde{\rho }_2  \sin^2{ {k} {x}})~,\label{energySoiton}\\
&\left< \mathcal{T}^{  x  x}\right> =\frac{r_0^3}{16\pi G L^4} (\tilde{\mathit{m}} +6 \tilde{w}_{1,3}+4 \tilde{\rho }_1 \tilde{\rho }_2  \sin^2{ {k} {x}})~, \\
&\left< \mathcal{T}^{  \chi  \chi}\right> = \frac{r_h^3}{16\pi G L^4} (-2 \tilde{m}- 6 \tilde{w}_{1,3} - 8 \tilde{\rho}_1 \tilde{\rho}_2 + 4 \tilde{\rho}_1 \tilde{\rho}_2 \sin^2{ {k} {x}})~.
\end{align}

Now let us consider Euclidean on-shell action, which can be obtained by analytic continuation of the total action $\mathcal{S}_{tot}$ in (\ref{bulk_action_2}). In order to get the on-shell action, it is useful to consider the following expression for the bulk part of the on-shell action: 
\begin{align}\label{Bulk_A}
\sqrt{-g}\mathcal{L}_{bulk} 
=-\bigg(\frac{2r U(r)}{L^4} e^{W_0(r)-W_0(\infty)+W_1(r)-W_1(\infty)} \bigg)' ~,
\end{align}
where we have used the equations of motion. Together with boundary counter terms in (\ref{Counter00}), one can see that the Euclidean on-shell action for black branes is given by 
\begin{align}\label{onShell}
\mathcal{S}_{on-shell}
=& \frac{r_h^3 }{16\pi G L^4}  \int_0^{1/T} d {\tau}_E  \int d^2 {x} \,   \left(-\tilde m -4 \tilde{\rho}_1 \tilde{\rho}_2 \sin^2{ {k} {x}} \right)\nonumber\\
=&\lim_{L_x \to \infty}\frac{r_h^3 L_x }{16\pi G L^4}  \int_0^{1/T} d {\tau}_E  \int d {y} \,  \left(-\tilde m -2 \tilde{\rho}_1 \tilde{\rho}_2  \right)~.
\end{align}
For the AdS soliton, one can also use this expression by replacing $r_h$ with $r_0$.

\subsection*{B. Equations of Motion and Regularity Condition}

Together with the  scaling (\ref{scaling00}), the equations of motion based on the ansatzs (\ref{BBrane00}) and (\ref{solition01}) are given by
\begin{align}\label{EOM}
u'&-\frac{2 \left(\rho ^2 e^{-2 w_1} \tilde{k}^2+\left(\rho ^2-3\right) \left(\rho ^2-1\right) \tilde{r}^2\right)}{\left(\rho ^2-1\right)^2 \tilde{r} \left(\tilde{r} w_1'+2\right)}-\frac{2 \left(\rho ^2-3\right) \tilde{r}^2 w_1'}{\left(\rho ^2-1\right) \left(\tilde{r} w_1'+2\right)}\nonumber\\\nonumber
&+\frac{u \left(2 \tilde{r}^2 \left(\rho '\right)^2+2 \left(\rho ^2-1\right)^2 \left(\tilde{r} w_1'+1\right){}^2\right)}{\left(\rho ^2-1\right)^2 \tilde{r} \left(\tilde{r} w_1'+2\right)}=0\\
w_1''&+\frac{2 \rho ^2 e^{-2 w_1} \tilde{k}^2}{\left(\rho ^2-1\right)^2 u \tilde{r}^2}+\frac{\left(\rho ^2-3\right) \tilde{r} w_1'}{\left(\rho ^2-1\right) u}+\frac{w_1'}{\tilde{r}}=0\nonumber\\
w_0'&+\frac{2 \rho ^2 e^{-2 w_1} \tilde{k}^2}{\left(\rho ^2-1\right)^2 u \tilde{r} \left(\tilde{r} w_1'+2\right)}+\frac{w_1' \left(\left(\rho ^2-3\right) \tilde{r}^2-\left(\rho ^2-1\right) u \left(\tilde{r} w_1'+1\right)\right)}{\left(\rho ^2-1\right) u \left(\tilde{r} w_1'+2\right)}\nonumber\\\nonumber
&-\frac{2 \tilde{r} \left(\rho '\right)^2}{\left(\rho ^2-1\right)^2 \left(\tilde{r} w_1'+2\right)}=0\\
\rho ''& +\frac{\rho ' \left(\left(\rho ^2-3\right) \tilde{r}^2+\left(\rho ^2-1\right) u\right)}{\left(\rho ^2-1\right) u \tilde{r}}-\frac{2 \rho  \left(\rho '\right)^2}{\rho ^2-1} +\frac{\rho  }{u}\left(\frac{\left(\rho ^2+1\right) e^{-2 w_1} \tilde{k}^2}{\left(\rho ^2-1\right) \tilde{r}^2}+2\right) =0~,
\end{align} 
where we used the scaling with $r_0$ instead of $r_h$ for the AdS Q-soliton. 
As we discussed earlier, these equations hold true for the AdS Q-soliton and the Q-lattice black brane. The shooting method has been taken to find the numerical solutions. This method considers difference equations approximately in terms of the discrete radial position $\hat{r}_i = 1 + \Delta\, (i-1)$ instead of the continuous $\tilde{r}$, where $i$ is a positive integer and $\Delta$ is the lattice spacing.

For a given initial condition set $\left\{\,u(1)=0, w_1(1), w_1'(1), w_0(1), \rho(1), \rho'(1)\,\right\}$, one can compute the set of data $\left\{ u'(1), w_1''(1), w_0'(1), \rho''(1) \right\}$ using the equations of motion  (\ref{EOM}). However, the second, third and forth equations of motion contain terms having $u$ in the denominators. These dangerous terms give divergent $w_1''(1), w_0'(1)$ and $\rho''(1)$, so the resultant solution is not a regular geometry. In order to avoid such a singular configuration, we have to expand the equations of motion near $\tilde{r}=1$. Then, one can isolate the dangerous terms and find constraints of the initial condition which produce finite $w_1''(1), w_0'(1)$ and $\rho''(1)$. The second and third equations of motion in (\ref{EOM}) share the same constraint given by the last regularity condition in (\ref{regul_con_BH}). The last equation of (\ref{EOM}) gives the initial condition constraint given by the second condition in (\ref{regul_con_BH}). By imposing the regularity condition (\ref{regul_con_BH}), one can continue the computation for $i\geq 2$ and regular solutions can be produced numerically.


\end{document}